\documentclass[manuscript]{acmart}

\usepackage{tabularx}
\usepackage{enumitem}
\usepackage{multirow}
\usepackage{subcaption}
\usepackage{soul}
\usepackage{colortbl}

\AtBeginDocument{%
  }

\copyrightyear{2025}
\acmYear{2025}
\setcopyright{cc}
\setcctype{by-nc-nd}
\acmConference[CHI '25]{CHI Conference on Human Factors in Computing Systems}{April 26-May 1, 2025}{Yokohama, Japan}
\acmBooktitle{CHI Conference on Human Factors in Computing Systems (CHI '25), April 26-May 1, 2025, Yokohama, Japan}
\acmDOI{10.1145/3706598.3713522}
\acmISBN{979-8-4007-1394-1/25/04}

\definecolor{blue20}{RGB}{208, 226, 255}
\definecolor{blue30}{RGB}{166, 200, 255}

\newcommand{\say}[1]{\textit{``#1''}}

\begin{document}

\title{Which Contributions Deserve Credit? Perceptions of Attribution in Human-AI Co-Creation}

\author{Jessica He} 
\email{jessicahe@ibm.com}
\orcid{0000-0003-2368-0099}
\affiliation{%
  \institution{IBM Research}
  \city{Seattle}
  \state{WA}
  \country{USA}
}

\author{Stephanie Houde}
\email{stephanie.houde@ibm.com}
\orcid{0000-0002-0246-2183}
\affiliation{%
  \institution{IBM Research}
  \city{Cambridge}
  \state{MA}
  \country{USA}
}

\author{Justin D. Weisz}
\email{jweisz@us.ibm.com}
\orcid{0000-0003-2228-2398}
\affiliation{%
  \institution{IBM Research}
  \city{Yorktown Heights}
  \state{NY}
  \country{USA}
}


\renewcommand{\shortauthors}{He et al.}

\begin{abstract}
    AI systems powered by large language models can act as capable assistants for writing and editing. In these tasks, the AI system acts as a \emph{co-creative partner}, making novel contributions to an artifact-under-creation alongside its human partner(s). One question that arises in these scenarios is the extent to which AI should be credited for its contributions. We examined knowledge workers' views of attribution through a survey study (N=155) and found that they assigned different levels of credit across different contribution types, amounts, and initiative. Compared to a human partner, we observed a consistent pattern in which AI was assigned less credit for equivalent contributions. Participants felt that disclosing AI involvement was important and used a variety of criteria to make attribution judgments, including the quality of contributions, personal values, and technology considerations. Our results motivate and inform new approaches for crediting AI contributions to co-created work.
\end{abstract}





\maketitle

\section{Introduction}

As large language models (LLMs) are incorporated into co-creative workflows in increasingly complex ways, one challenge that arises is determining how to delineate authorship. Although there are well-established standards for crediting contributors when writing with human collaborators, requirements for crediting the use of AI\footnote{Our use of ``AI'' in this paper is to succinctly refer to systems or applications that are powered by generative AI technologies; e.g. ``AI systems'' or ``AI-infused applications.'' It is not intended to anthropomorphize the underlying technology, even when we reference an ``AI partner.''} are nascent and often simplistic. Publishers, professional groups, and governments have begun to define policies on crediting AI. These policies tend to follow a one-size-fits-all approach in which \emph{any} AI involvement is required to be acknowledged, with few guidelines on how to distinguish between different kinds of AI contributions~\cite{acm2023policy, arxiv, california2024ai, cooley2024, eu2023ai, icmje, nature, wiley2023generative, zirpoli2023generative}. For example, the United States Patent and Trade Office recently issued guidance that states, ``those involved in patent proceedings have a duty to disclose all information—including on the use of AI tools by inventors, parties, and practitioners—that is material to patentability''~\cite[Section A]{uspto24guidance}.

In contrast, prior research has found that people's perceptions of AI authorship in co-creation are nuanced. These perceptions depend on various dimensions, including the type of contribution made by the AI, how much each party contributed, and who led the co-creative effort~\cite{he2024ai, lee2022coauthor, rezwana2023user, xu2024makes}. We build on this body of work and re-examine attribution policies by understanding how AI should be attributed across different co-creative scenarios from the perspective of people authoring written content with AI. Specifically, we seek to understand peoples' views on the extent to which AI co-creative partners deserve credit for different contributions, determine how these views compare to crediting a human partner, and identify factors salient when attribution decisions are made. Given the recent push from businesses to integrate generative AI into knowledge work to boost productivity~\cite{alavi2023} and the increasing use of LLMs in such work~\cite{brachman2024knowledge, liang2024mapping}, we focus on knowledge workers in this study. We conducted a scenario-based survey with 155 knowledge workers who rated how much authorship credit should be given to either a human or AI partner for different contribution types, contribution amounts, and levels of initiative. Our work makes the following contributions:

\begin{enumerate}
    \item We demonstrate how a binary approach to AI attribution is insufficient and instead requires more granularity. Through an analysis of participants' authorship credit assignments to an AI partner across multiple scenarios, we found that different types and amounts of contributions, along with different forms of human and AI initiative, warranted different levels of authorship credit.
    \item We identify a set of factors that impact people's decisions on how much authorship credit an AI partner deserves. These factors were derived through a thematic analysis of participants' comments on how they made attribution decisions.
    \item We find a consistent pattern in which AI partners were assigned less credit compared to human partners for equivalent contributions, signaling a need for AI-specific attribution frameworks.
    \item We propose new design strategies for capturing and communicating nuances in AI contribution to stimulate discussion on future AI attribution frameworks.
\end{enumerate}

Our findings indicate that AI attribution is not one-size-fits-all, and that there is a need for new attribution approaches that not only acknowledge the involvement of AI in co-created work, but also show \emph{how} AI contributed. As AI transparency requirements take effect across industries and geographies~\cite{california2024ai, eu2023ai}, we anticipate our work to help inform approaches for meeting those requirements in ways that reflect the views of content creators.

\section{Related Work}

We outline three areas relevant to our study of AI attribution in co-creative writing scenarios. First, we provide a brief overview of the emerging field of human-AI co-creation, in which generative AI models and human users both manipulate artifacts-under-creation. We then discuss prior studies that examined people's feelings of ownership over works co-created with AI. Finally, we examine the landscape of existing attribution frameworks to motivate the need for frameworks specific to AI co-creative partners.

\subsection{Co-creation with generative AI}

Generative AI is often used for co-creation across many content modalities, including text~\cite{wan2024felt, yang2022ai, ding2023mapping, lee2024design}, images~\cite{turchi2023human, fan2024contextcam}, audio~\cite{cao2023comprehensive, ning2024mimosa, louie2020novice}, and video~\cite{wang2024podreels, wang2024reelframer} due to its ability to produce high-fidelity works from natural language descriptions. One common co-creative task is writing~\cite{lee2024design}. Advancements in the ability for LLMs to produce fluent text has led to a rapid proliferation of AI-powered writing assistants, such as Copilot for Microsoft 365\footnote{Copilot for Microsoft 365: \url{https://www.microsoft.com/en-us/microsoft-365/enterprise/copilot-for-microsoft-365}} and Grammarly's AI Writing Assistant\footnote{Grammarly AI Writing Assistant: \url{https://www.grammarly.com/ai-writing-assistant}}. Researchers have explored the use of LLMs to support writing in many domains, including fiction~\cite{calderwood2020novelists, yang2022ai, zhao2023more}, blog posts~\cite{radensky2024let}, social media posts~\cite{lu2024corporate}, song lyrics~\cite{huang2020ai}, comedy~\cite{mirowski2024robot}, and more. \citet{lee2024design} mapped out dimensions within the design space for intelligent writing assistants and identified a wide range of writing contexts, purposes, and stages that they can support. Their work highlights several writing tasks in which AI can contribute, including idea generation, planning, drafting, and revision~\cite{lee2024design}. As such, co-writing with AI assistants can occur through a variety of workflows: a user might use AI-generated ideas as inspiration, write their own draft and ask AI to revise it, ask AI to write a full draft then review it, or anywhere in between. 

Studies of AI-assisted writing have indicated that LLMs can improve various facets of writing, including the creativity, confidence, and efficiency of writers~\cite{cardon2023challenges, doshi2024generative, li2024value}. Studies with students have similarly found that AI assistance can improve their writing skills~\cite{amyatun2023can, song2023enhancing}. In one study on using LLMs to write social media posts, \citet{long2023tweetorial} found that LLMs reduced the mental demand and effort required of writers by supporting brainstorming, information gathering, and revision, hence improving their confidence and satisfaction with their work. In the workforce, knowledge workers have begun using LLMs for writing tasks such as ideation, improving text, and creating a first draft~\cite{brachman2024knowledge}.

The incorporation of AI assistance into writing workflows raises new questions regarding the need for transparency and explainability~\cite{sun2022investigating}. The variable nature of generated outputs -- what \citet{weisz2024design} call ``generative variability'' -- along with the propensity for LLMs to produce factually incorrect yet plausible-sounding text ~\cite{hicks2024chatgpt}, has led researchers to explore techniques for mitigating overreliance on generated texts: attributing sources of information~\cite{do2024facilitating}, identifying and drawing attention to model uncertainties~\cite{kim2024m, weisz2021perfection}, and predicting the likelihood of requiring human edits~\cite{vasconcelos2023generation}. Tools such as iA Writer\footnote{iA Writer for Mac: \url{https://ia.net/writer/support/editor/authorship}} distinguish AI-generated text via different text styles. Technologies such as GLTR\footnote{GLTR: \url{http://gltr.io}} and Radar~\cite{hu2023radar} detect AI-generated text in writings of unknown origins~\cite{gehrmann2019gltr}. Although these studies and tools make important strides in improving transparency in LLM use, there remains a gap in understanding people's transparency needs surrounding the authorship of co-created content -- in essence, much research has focused on understanding \emph{how} an LLM produced a text, but less emphasis has been placed on communicating \emph{what was produced} and \emph{how it was used}. Given the prevalence of AI-assisted writing tools and the growing complexity of human-AI co-writing workflows, there is a need to better understand people's perceptions of attribution and desires for transparency in this rapidly-evolving space.

\subsection{User perceptions of ownership in co-created work}
\label{sec:related-work-perceptions}

\citet{lee2024design} identified ownership as an important dimension of AI-assisted writing, defining it as a ``user’s sense of ownership or authenticity over the written artifact when using the writing assistant.'' As co-writing workflows become more complex and pervasive, the line between AI and human ownership over co-created work becomes blurrier: who is an author and how should each party be credited? \citet{yeh2024ghostwriter} found mixed perceptions of ownership over works co-written with an AI writing tool. \citet{he2024ai} found that, ``Participants mostly felt that ownership was shared between the human and the AI'' in an AI-assisted brainstorming scenario. \citet{draxler2024ai} explored the relationship between sense of ownership and declaration of authorship in the context of co-writing postcards for a friend. They found an ``AI ghostwriter effect'' in which people were less likely to declare an AI's involvement compared to the involvement of another person, even though they did not consider themselves to be the owners or authors of AI-generated writing. Even so, 43.3\% of participants in one of their studies felt that disclosing AI contributions should be mandatory for ethical, transparency, and self-protection reasons, and 61.5\% of participants in their second study felt that denoting AI involvement should be mandatory~\cite{draxler2024ai}.

Work in the HCI literature has also begun to identify factors that affect people's sense of ownership over work produced with AI assistance. Some of these factors regard characteristics of the contributions themselves, such as the types of contributions made~\cite{rezwana2023user, he2024ai, wan2024coco, xu2024makes} and the amount of material contributed~\cite{lee2022coauthor}. Other factors focus on the process by which contributions were made. For example, people's feelings of ownership increase when performing the work of ``critical integration'''~\cite{sarkar2023exploring} or with a greater sense of control over the work~\cite{draxler2024ai, louie2020novice, xu2024makes}. The role of AI in the writing process can also influence ownership perceptions~\cite{draxler2024ai, rezwana2023user, gero2019metaphoria, xu2024makes}. \citet{rezwana2023user} reported the role of AI as a tool vs. an independent entity as a salient consideration, and \citet{gero2019metaphoria} found that a cognitive offloading tool that was used ``like a calculator for words'' raised fewer concerns over ownership compared to a co-creative partner who made more substantial contributions. In the same vein, several works have explored interventions to prevent the loss of human ownership over co-created work. Techniques such as writing longer prompts~\cite{joshi2024writing}, using AI for ideation or a starting point~\cite{biermann2022tool}, and viewing AI contributions in a sidebar rather than within the text~\cite{kim2024towards} have all been found to increase people's sense of ownership, as well as designing AI assistants to be more interactive, transparent, and personalized~\cite{gero2019metaphoria, neate2019empowering}.

One key process factor that impacts feelings of ownership is initiative: whether a contribution was initiated by the human or AI partner~\cite{rezwana2023user}, and whether that contribution was made via ``direct generation'' or ``indirect suggestions''~\cite{wan2024coco}. In our study, we sampled three primary contribution dimensions -- contribution type, contribution amount, and initiative -- because they cover characteristics of the contributions themselves and the process by which they were generated. We summarize these dimensions in Table~\ref{tab:contribution-dimensions}.

\begin{table}[htp]
    \centering
    \small
    \begin{tabularx}{\linewidth}{p{1.5cm}XX}
        \toprule
        \textbf{Dimension} & \textbf{Definition} & \textbf{Literature} \\
        \midrule
        Type of\newline contribution & The kind of content contributed by the writing partner. & \citet{he2024ai, rezwana2023user, wan2024coco, xu2024makes} \\
        \midrule
        Amount of contribution & The quantity of content contributed by the writing partner. & \citet{lee2022coauthor}  \\
        \midrule
        Initiative & Whether the writing partner made a proactive or reactive contribution and whether they contributed actual content or recommendations. & \citet{lee2024design, rezwana2023user, wan2024coco}  \\
        \bottomrule
    \end{tabularx}
    \caption{Summary of dimensions affecting perceptions of ownership in co-creation that we used to structure our study, identified from prior HCI and co-creative literature.}
    \Description{Summary of dimensions affecting perceptions of ownership in co-creation that we used to structure our study, identified from prior HCI and co-creative literature. Each row contains a dimension (type of contribution, amount of contribution, and initiative), its definition, and a list of prior literature that identified it.}
    \label{tab:contribution-dimensions}
\end{table}

These studies provide valuable insights into people's feelings of ownership over co-created works, which inform the focus of our paper: \emph{attribution}. Attribution is the visible recognition provided to the author of a work. Attribution differs from ownership in two key ways. From a legal perspective, ownership provides certain rights and accountability, such as being able to revive work that is no longer being disseminated~\cite{van2016authors}. From a psychological perspective, ownership refers to a ``sense of possession over the target''~\cite{xu2024makes}. In both cases, ownership does not entail a requirement for attribution. The relationship between attribution and authorship is more complex. In many cases, authors are attributed as contributors to their work. However, it is not always required for an author to be attributed (e.g., ghostwriting), or to be attributed under their own name (e.g., pen names).

In this work, we focus on understanding how people believe authorship should be attributed across the many potential types of co-creative workflows. What type of credit should AI receive for different levels of involvement in the co-creative process? While \citet{draxler2024ai} studied similar questions, we sought to conduct a more granular analysis across different natures of contribution, writing contexts, and types of authorship credit.

\subsection{Attribution practices}
\label{sec:attribution-practices}

When writing collaboratively with other people, there are well-established standards for crediting contributions. For example, the ACM's criteria state that authors must ``make substantial intellectual contributions to some components of the original Work'' and ``take full responsibility for all content in the published Works''~\cite{acm2023policy}. The CRediT taxonomy, used by over 120 journals as of 2019, identifies contribution types that can be specified in authorship statements~\cite{brand2015beyond}. Researchers have also explored interactive crediting~\cite{bd2016solving} and creative representations of author order~\cite{demaine2023every}. In the editing domain, different roles capture different contribution responsibilities. For example, developmental editors review the big picture and ensure conceptual alignment, whereas copy editors revise mechanical aspects such as grammar, spelling, and wording~\cite{reeder2016three}. Editors are often acknowledged but not typically listed as authors in published work.

The extent to which these standards translate to human-AI co-creation is an open question. Model providers have begun defining guidelines for attribution of works created with their systems. For example, OpenAI states that when co-authoring with their API, ``published content is attributed to your name or company,'' and requires that ``the role of AI in formulating the content is clearly disclosed''~\cite{openai2022sharing}. Generative AI applications are taking a similar approach. One example is ``Draft One,''~\footnote{Draft One. \url{https://www.axon.com/products/draft-one}} which uses an LLM to assist police officers in writing reports and requires officers to check a box indicating AI usage~\cite{murphy2024police}. Similarly, many journals and publishing companies require the use of AI to be acknowledged, such as in a Methods or Acknowledgments section. However, they maintain that AI cannot credited as an author~\cite{acm2023policy, arxiv, elsevier, icmje, wiley2023generative}. For example, Nature's guidelines state, ``Large Language Models...do not currently satisfy our authorship criteria. Notably an attribution of authorship carries with it accountability for the work, which cannot be effectively applied to LLMs'' \cite{nature}. The language of ``currently satisfy'' indicates that their requirements may evolve as generative technology advances. 

Legal guidelines have also begun to emerge around AI authorship. In the United States, the Copyright Office has declared that AI cannot be recognized as an owner of creative work -- only human portions of co-created work are protected by US copyright~\cite{zirpoli2023generative}. The Copyright Office has stated that they may issue additional guidance as the technology evolves~\cite{zirpoli2023generative}. In contrast, two court decisions in China have indicated that AI-generated content is eligible for copyright protection by a user or platform developer if it is an ``original intellectual achievement''~\cite{cooley2024}. Similarly, in the United Kingdom, copyright law provides protection for ``computer-generated works'' because whoever made the ``arrangements necessary for the creation of the work'' is designated the owner~\cite{cooley2024}. In the EU AI Act, legal guidance on AI authorship is notably missing~\cite{cooley2024}. Transparency requirements are also emerging; the EU AI Act~\cite{eu2023ai} and the California AI Transparency Act~\cite{california2024ai} both require disclosure of AI-generated content, but neither specifies how to do so. The variability in authorship guidelines across different geographies, along with the lack of concrete guidance on how to provide transparency, reflect the uncertain and evolving nature of human-AI authorship.

Although emerging policies have begun to define requirements for AI authorship and attribution, they often take a one-size-fits-all approach that does not account for different kinds of contributions in co-creative workflows, nor do they account for the attribution preferences of those creating the content. As \citet{sarkar2023exploring} contends, ``creativity is ultimately defined by communities of creators and receivers.'' 
A creator-centric \emph{how} of AI attribution remains an open question -- while algorithms for identifying and watermarking AI-generated content offer useful technological approaches~\cite{dathathri2024scalable, fernandez2023stable, hu2023radar}, they may not encompass the extent and nuance of creators' attribution needs. We aim to understand a more granular set of requirements for AI attribution in co-creativity from the perspective of creators: how much authorship credit should be given to AI for different kinds of contributions, and how should their perceptions translate into guidelines for crediting AI?

\section{Study of Attribution Perceptions}

Our work explores three research questions on people's views and decision processes in attributing co-created work.

\begin{itemize}
    \item \textbf{RQ1}. What are people's views on attribution when working with an AI partner across different co-creative writing scenarios?
    \item \textbf{RQ2}. How do people's views on attribution differ between human and AI partners?
    \item \textbf{RQ3}. How do people make attribution decisions?
\end{itemize}

\subsection{Design}
\label{section:study-design}

We conducted a survey study to assess people's perceptions of attribution across different scenarios. Our study used a 2 $\times$ 3 factorial design with \emph{writing partner} (human or AI) and \emph{writing context} (academic, professional, or technical) as between-subjects factors. Each participant received one of six unique scenario variations (Table \ref{tab:study-conditions}), comprised of one partner and one context, which were assigned randomly. The full text for each scenario can be found in Appendix~\ref{appendix:scenarios}.

To begin, participants were provided with context for the scenario: they were an employee of a large, international technology company and were authoring a written text as part of their work with the help of a partner. They then responded to 18 questions about their perceptions of attribution across different dimensions, 4 questions about how they made their attribution decisions, and 2 questions about their use of generative AI.

\subsubsection{Writing partner}
Given our goal of comparing how people felt about attribution for AI and human partners (RQ2), we randomly assigned participants to a scenario in which their writing partner was either an ``AI system'' or a human ``colleague.'' No further details of their partner beyond these descriptions were provided.

\subsubsection{Writing context}

Writing takes many different forms and serves many different purposes. \citet{lee2024design} identified a set of six writing contexts relevant to intelligent writing assistants: academic, creative, journalistic, technical, professional, and personal. Given our focus on writing that occurs in workplace settings, and our desire to limit the combinatorial complexity of our study, we examined three representative contexts: academic, technical, and professional. We did not formulate any particular hypotheses about differences in attribution across these contexts; rather, we included multiple contexts to increase the generalizability of our study.

\subsubsection{Dimensions of co-creative attribution}
\label{sec:study-design-dimensions}

To address our goal of understanding attribution perceptions across different co-creative scenarios (RQ1), we identified three dimensions from prior work that may impact people's authorship perceptions in co-creation: type of contribution, amount of contribution, and initiative. Within each dimension, we identified different \emph{natures of contribution} by considering phases of writing~\cite{flower1981cognitive, lee2022coauthor, wan2024coco}, co-creative actions~\cite{rezwana2023designing, spoto2017mici, wan2024coco}, and the roles of writers and editors in human-human~\cite{burrough2013defining, lowry2004building, posner1992people} and human-AI~\cite{gero2019metaphoria, ippolito2022creative, rezwana2023user, xu2024makes} writing (previously discussed in Section~\ref{sec:related-work-perceptions}).

For contribution type, we synthesized a set of representative examples identified from prior literature on collaborative writing. \citet{burrough2013defining} and \citet{sarkar2023exploring} both differentiate between ``superficial'' or ``form'' edits that affect stylistic choices and ``deep'' or ``content'' edits that alter meaning or ideas.
Hence, in identifying types of contributions, we made sure to include both contributions of \emph{form} (spelling \& grammar~\cite{burrough2013defining, wan2024coco}, organization \& structure~\cite{burrough2013defining, wan2024coco}, readability \& clarity~\cite{burrough2013defining, wan2024coco}, tone \& style~\cite{burrough2013defining, ippolito2022creative}) and contributions of \emph{content} (fact check~\cite{burrough2013defining}, elaborate on ideas~\cite{ippolito2022creative, rezwana2023designing}, add new ideas~\cite{ippolito2022creative, lee2022coauthor, lee2024design, rezwana2023designing, spoto2017mici}, narrow scope~\cite{burrough2013defining}, synthesize information~\cite{posner1992people, wan2024coco}). In some instances, the literature explicitly described the contribution type as form or content; in others, we categorized it based on \citet{burrough2013defining}'s and \citet{sarkar2023exploring}'s definitions. For contribution amount, we considered a gradation ranging from no writing to all writing, with points in between covering different proportions. For initiative, we explored several tools and ideas regarding how proactively a co-creative partner might act, and we recognized that there were different ways of being proactive. In one form of proactivity, the partner might contribute on its own compared to a partner that only acts when asked (i.e., the ``respond'' vs. ``initiate'' dichotomy of \citet{muller2022frameworks} and the ``user-initiated'' vs. ``system-initiated'' dimensions of \citet{lee2024design}). In another form of proactivity, the partner might directly create content for an artifact, compared to a partner that only makes recommendations (e.g., an LLM that grades essays and provides feedback~\cite{stahl2024exploring})~\cite{lee2024design, wan2024coco}. Hence, we crossed both interpretations of proactivity to form four variants. We recognize that people's experiences with proactive AI may be limited -- although some degree of proactivity exists in the form of LLM-based autocompletion tools (e.g., Apple Intelligence\footnote{Apple Intelligence: \url{https://www.apple.com/apple-intelligence/}.} and Microsoft Copilot\footnote{Microsoft Copilot: \url{https://www.microsoft.com/en-us/microsoft-365/copilot/}.}), tools that exhibit more extensive levels proactivity, such as producing complete artifacts without human input, is not yet widespread. However, we anticipated that participants could effectively speculate on such scenarios based on their experiences with both existing AI autocompletion tools and human collaborators, and we found those scenarios to be valuable to study given the recent emergence of agentic AI technology that can enable such levels of proactivity~\cite{zhang2023proagent, xi2023rise} and recent HCI interest in studying proactive AI agents~\cite{chen2024need, muller2024group}.

We show our taxonomy of contributions in Table~\ref{tab:dimensions}. To validate our taxonomy, we conducted a small pilot test in which we asked four people to fill out our survey while thinking aloud~\cite{van1994think}. This pilot resulted in the addition of missing types and amounts of contribution and the collapsing of redundant parts of our taxonomy.

\begin{table*}[htp]
    \centering
    \small
    \begin{tabularx}{\linewidth}{>{\raggedright\arraybackslash}p{4.5cm}X}
        \toprule
        \textbf{Contribution} & \textbf{Description} \\
        \midrule
        \textit{Type} & \textit{The kind of content contributed by the writing partner.} \\
        Spelling \& grammar       & Your partner makes spelling and grammar corrections. \\
        Organization \& structure & Your partner makes revisions to organization and structure. \\
        Readability \& clarity    & Your partner makes improvements to readability and clarity. \\
        Tone \& style             & Your partner makes revisions to tone and style to ensure the writing is suitable for the target audience. \\
        Fact check                & Your partner identifies outdated or incorrect information to correct. \\
        Elaborate on ideas        & Your partner elaborates on existing ideas. \\
        New ideas                 & Your partner introduces new ideas. \\
        Narrow scope              & Your partner narrows the scope by removing less important ideas. \\
        Synthesize information    & Your partner synthesizes your notes and ideas into cohesive writing. \\
        \midrule
        \textit{Amount} & \textit{The quantity of content contributed by the writing partner.} \\ 
        No writing      & You write the text on your own, without assistance from a partner.\\
        A few sentences & You write most of the text and ask your partner to write a few sentences. \\
        Equal writing   & You write a few paragraphs and ask your partner to write a few other paragraphs. \\
        Most writing    & You write a few sentences and ask your partner to write the rest. \\
        All writing     & You ask your partner to write the full text. \\
        \midrule
        \textit{Initiative} & \textit{Whether the writing partner made a proactive or reactive contribution and whether they contributed actual content or recommendations.} \\
        Proactively writes complete text  & Your partner proactively suggests complete written text without being asked. \\
        Asked to write complete text      & You ask your partner to write the complete text.  \\
        Proactively makes recommendations & Your partner proactively provides ideas or feedback without being asked. \\
        Asked to make recommendations     & You ask your partner for ideas or feedback to incorporate into the work. \\
        \bottomrule
    \end{tabularx}
    \caption{Descriptions of different natures of contribution across the dimensions of contribution type, amount, and initiative. The full texts of the descriptions used in the study are shown in Appendix~\ref{appendix:survey-instrument}.}
    \Description{Descriptions of different natures of contribution across the dimensions of contribution type, amount, and initiative. The full texts of the descriptions used in the study are shown in Appendix~\ref{appendix:survey-instrument}.}
    \label{tab:dimensions}
\end{table*}

For each nature of contribution, participants were asked to indicate what kind of authorship credit their partner should receive. Ratings were made on a 7-point scale\footnote{In the response options, ``[partner]'' was replaced with either ``AI'' or ``your colleague'' depending on the participant's condition.}: ``You are the sole author,'' ``You are the primary author; [partner] is acknowledged but not as an author,'' ``You are the primary author; [partner] is the secondary author,'' ``You and [partner] have equal authorship,'' ``[Partner] is the primary author; you are the secondary author,'' ``[Partner] is the primary author; you are acknowledged but not as an author,'' ``[Partner] is the sole author,'' and ``Unsure.'' We provide additional details on how we used this scale in our analysis in Table~\ref{tab:credit-assignment-scale}, and we provide the full survey instrument in Appendix~\ref{appendix:survey-instrument}. As some of the contributions we studied (e.g., new ideas, recommendations) may only have shaped the final artifact in invisible ways, our instructions to participants were to assume that any contributions made were included in the final artifact.

We also asked participants questions about how they made their attribution decisions. They rated the importance of each dimension -- contribution type, amount, and initiative -- on a 5-point scale: ``Not at all important,'' ``Not very important,'' ``Neither important nor unimportant,'' ``Important,'' and ``Very important.'' They also optionally described, in their own words, how they determined the appropriate type of attribution and what factors affected their decisions. We concluded the survey with questions about how frequently participants used generative AI and in what contexts.

\subsection{Participants}
\label{sec:participants}

We conducted our survey within our organization, a large, multi-national technology company. We recruited knowledge workers from internal Slack channels for employees across different business units, geographic regions, and professional interests to obtain a diverse sample.

We received 184 responses to our survey. Of these, we filtered out 29 for several reasons. First, we sought participants familiar with using generative AI since they would be more likely to understand the nuances of AI attribution; hence, we filtered out responses that indicated no generative AI experience. Second, we treated one of the contribution amounts as an attention check to filter out low-quality responses: for the question, \emph{``You write the artifact on your own, without [partner's] assistance,''} we removed responses that were not \emph{``You are the sole author.''} Finally, we filtered out cases in which a respondent selected ``Unsure'' across all natures of contribution, as they did not provide any useful signal.

After filtering, we were left with a final sample of 155 responses. Due to filtering, the number of participants within each of the six conditions was slightly uneven (Table~\ref{tab:study-conditions}), although the overall split of participants assigned to the human (N=78) and AI (N=77) conditions remained even. In describing our results, we refer to participants as \texttt{Pxx-\{H,AI\}} to indicate the participant's ID and whether they were in the human or AI partner condition.

Over 63\% of our participants were active users of generative AI applications on a daily (25.2\%) or weekly (38.1\%) basis; the rest reported usage frequencies of either monthly (19.3\%) or less frequently (17.4\%). They reported using generative AI for a variety of purposes, including personal use (71.6\%), school or academics (21.9\%), and work (59.4\%). Additionally, almost a third of participants (31.6\%) reported developing, designing, or studying generative AI as part of their job, with 10.3\% reporting that they were involved in the training or tuning of generative AI models.

Our participants were drawn from a variety of job roles, including software development \& support (24.5\%), consulting (12.9\%), marketing \& communications (9.7\%), management (6.5\%), design (6.5\%), and a long tail of other knowledge work roles (39.9\%). They were located in a variety of regions, with 63.2\% in the Americas, 19.4\% in Europe, the Middle East, or Africa (EMEA), and 15.5\% in Asia-Pacific (APAC); 1.9\% of participants did not disclose their location.

\begin{table}[htp]
    \centering
    \small
    \begin{tabularx}{\linewidth}{p{2cm}XXX}
        \toprule
         & \textbf{Academic context}  & \textbf{Professional context} & \textbf{Technical context} \\
        \midrule
        \textbf{Human partner} & Academic + Human (N=24) & Professional + Human (N=29) & Technical + Human (N=25) \\
        \midrule
        \textbf{AI partner}    & Academic + AI (N=25) & Professional + AI (N=28) & Technical + AI (N=24)  \\
        \bottomrule
    \end{tabularx}
    \caption{Study conditions and the number of participants within each condition. 78 participants were assigned to a scenario with a human partner and 77 were assigned to a scenario with an AI partner.}
    \Description{Study conditions and the number of participants within each condition. 78 participants were assigned to a scenario with a human partner and 77 were assigned to a scenario with an AI partner.}
    \label{tab:study-conditions}
\end{table}

Our survey took approximately 15 minutes to complete. To incentivize participation, we conducted a raffle and awarded 5 participants each the equivalent of \$25 USD.

\subsubsection{Ethics statement}
At-scale survey research conducted within our institution is subject to review and approval by an internal committee. Our study was approved by this committee, subject to restrictions regarding the collection of demographic information. Specifically, we were unable to collect participants' age or gender identity, we adhered to specified options to collect geography and job role, and we did not collect any personally-identifiable information (PII). All survey responses were anonymous.

\section{Results}

\subsection{Analysis}
Our study is primarily concerned with understanding how different natures of co-creative contribution impact a single outcome measure: authorship credit assignment. We measured this construct on a 7-point scale, where each point represented different degrees of credit assigned to the self and co-creative partner, which stemmed from common publishing practices (e.g.~\cite{acm2023policy, epa2024authorship, harvardauthorship}). For analysis, we convert the categorical scale points to numeric scores of [-3, +3], with ``Equal'' centered on 0. Table~\ref{tab:credit-assignment-scale} describes the scale points, their interpretation, and their numerical mappings. We refer to points on the categorical scale as \emph{ratings} and points on the numerical scale as \emph{scores}. 

\begin{table*}[htp]
    \centering
    \small
    \begin{tabularx}{\linewidth}{l>{\raggedright\arraybackslash}X>{\raggedright\arraybackslash}Xp{1cm}p{1.5cm}}
        \toprule
        \textbf{Categorical scale point} & \textbf{Self's credit} & \textbf{Partner's credit} & \textbf{Score} & \textbf{Range} \\
        \midrule
        Sole/n               & Self is the sole author and has complete authorship credit 
                             & Partner has no authorship credit 
                             & -3 & $[-3.0, -2.5)$ \\
        Primary/Acknowledged & Self is the primary author 
                             & Partner is acknowledged as a contributor 
                             & -2 & $[-2.5, -1.5)$\\
        Primary/Secondary    & Self is the primary author 
                             & Partner is a secondary author 
                             & -1 & $[-1.5,-0.5)$\\
        Equal                & Self is a co-author 
                             & Partner is a co-author 
                             & 0 & $[-0.5, 0.5)$ \\
        Secondary/Primary    & Self is a secondary author 
                             & Partner is the primary author 
                             & +1 & $[0.5, 1.5)$\\
        Acknowledged/Primary & Self is acknowledged as a contributor 
                             & Partner is the primary author 
                             & +2 & $[1.5, 2.5)$ \\
        n/Sole               & Self has no authorship credit 
                             & Partner is the sole author and has complete authorship credit 
                             & +3 & $[2.5, 3.0]$ \\
        \bottomrule
    \end{tabularx}
    \caption{Scale of attribution levels used in our analysis. Categorical scale points correspond to \textit{Self credit / Partner credit}. Participants also had the option to respond ``Unsure''; these ratings were treated as missing values in our analysis. The actual text used to describe the scale points is shown in Appendix~\ref{appendix:survey-instrument}. Numerical ranges are used to convert authorship credit scores to their categorical equivalents.}
    \Description{Scale of attribution levels used in our analysis. Categorical scale points correspond to Self credit / Partner credit. Participants also had the option to respond ``Unsure''; these ratings were treated as missing values in our analysis. The actual text used to describe the scale points is shown in Appendix~\ref{appendix:survey-instrument}. Numerical ranges are used to convert authorship credit scores to their categorical equivalents.}
    \label{tab:credit-assignment-scale}
\end{table*}

In analyzing authorship credit scores, we first used Shapiro-Wilk's test~\cite{shapiro1965analysis} to determine whether the data were normally distributed. They were not; authorship credit scores tended to be skewed toward the negative side of the scale, indicating a bias toward self-credit. This bias may have stemmed from the second-person framing of our scenarios: \emph{``As part of \underline{your} work, \underline{you're} writing...''} Therefore, to determine whether differences observed between different conditions (e.g. human vs. AI) were significant, we use nonparameteric tests. Specifically, we use the Wilcoxon rank-sum test\footnote{This test is also known as the Mann-Whitney $U$ test~\cite{mann1947test}.} to make pairwise comparisons between each writing partner; as our research questions do not specifically predict differences amongst different writing contexts, we omit this variable from our analysis\footnote{To ensure that writing context was not a significant factor, we conducted three ANOVA tests on authorship credit scores, one for each contribution dimension. Each model included writing context, writing partner, and their interaction as independent variables. Writing context was not a significant predictor of authorship credit scores across types of contribution (F(2,1379) = 0.40, p = n.s.), amounts of contribution (F(2,761) = .48, p = n.s.), or levels of initiative (F(2, 597) = .34, p = n.s.).}. We report the Wilcoxon rank-sum statistic as $W$ with an effect size of $r$\footnote{We interpret effect sizes $r < 0.3$ as small, $0.3 \leq r < 0.5$ as moderate, and $r \geq 0.5$ as large~\cite{cohen2013statistical}.}.

To analyze open-ended responses, we conducted a reflexive thematic analysis~\cite{braun2019reflecting, thomas2006general}. Through multiple rounds of discussion and iteration, two authors collaboratively labeled participants' comments, constructed a set of themes, and identified quotes that provide further insight into the quantitative results.

\subsection{Overview of authorship credit assignment trends}
\label{sec:overview-trends}

\begin{figure*}
    \centering
    \begin{subfigure}{0.33\textwidth}
        \includegraphics[width=\textwidth]{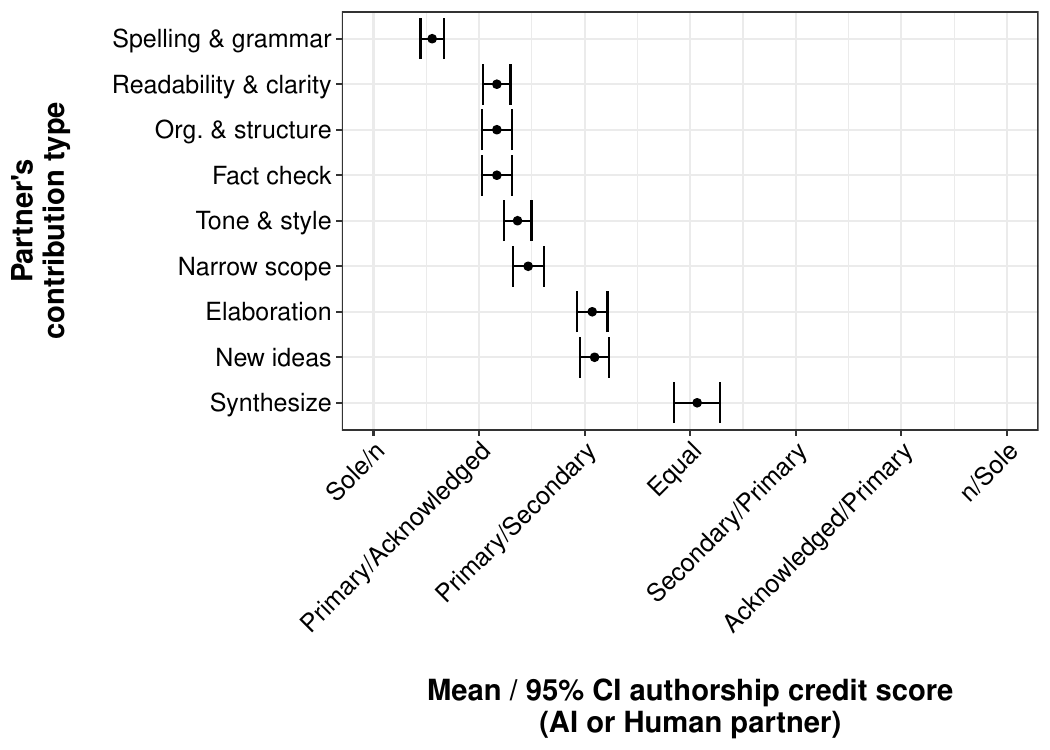}
        \caption{Contribution type}
    \end{subfigure}
    \begin{subfigure}{0.33\textwidth}
        \includegraphics[width=\textwidth]{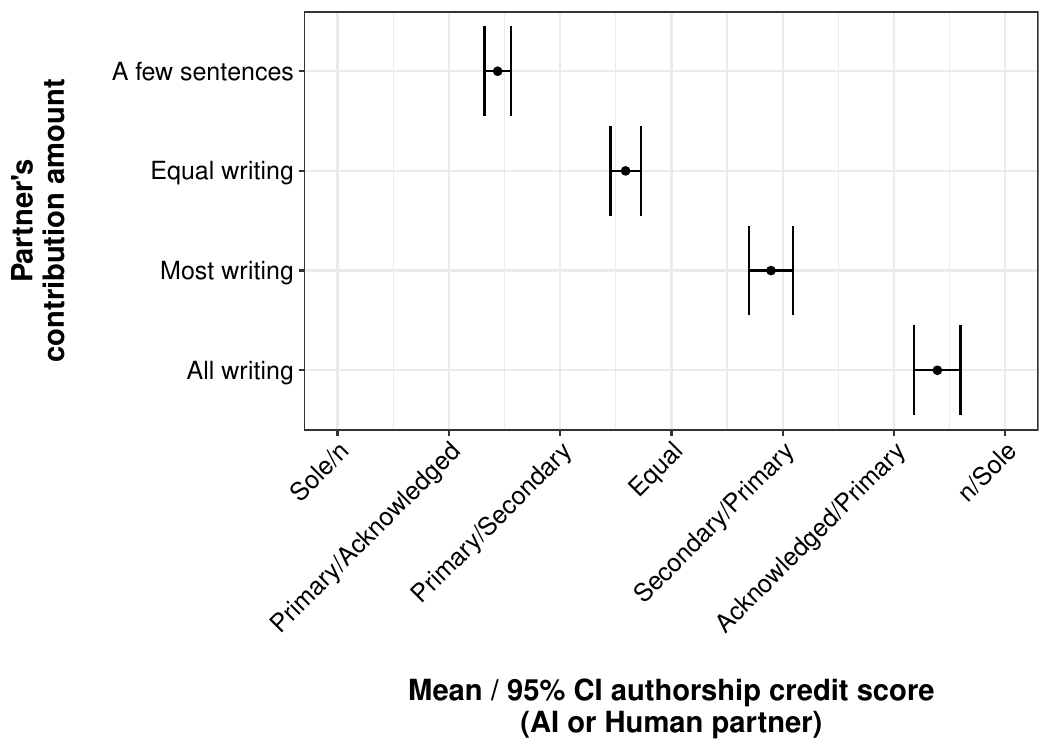}
        \caption{Contribution amount}
    \end{subfigure}
    \begin{subfigure}{0.33\textwidth}
        \includegraphics[width=\textwidth]{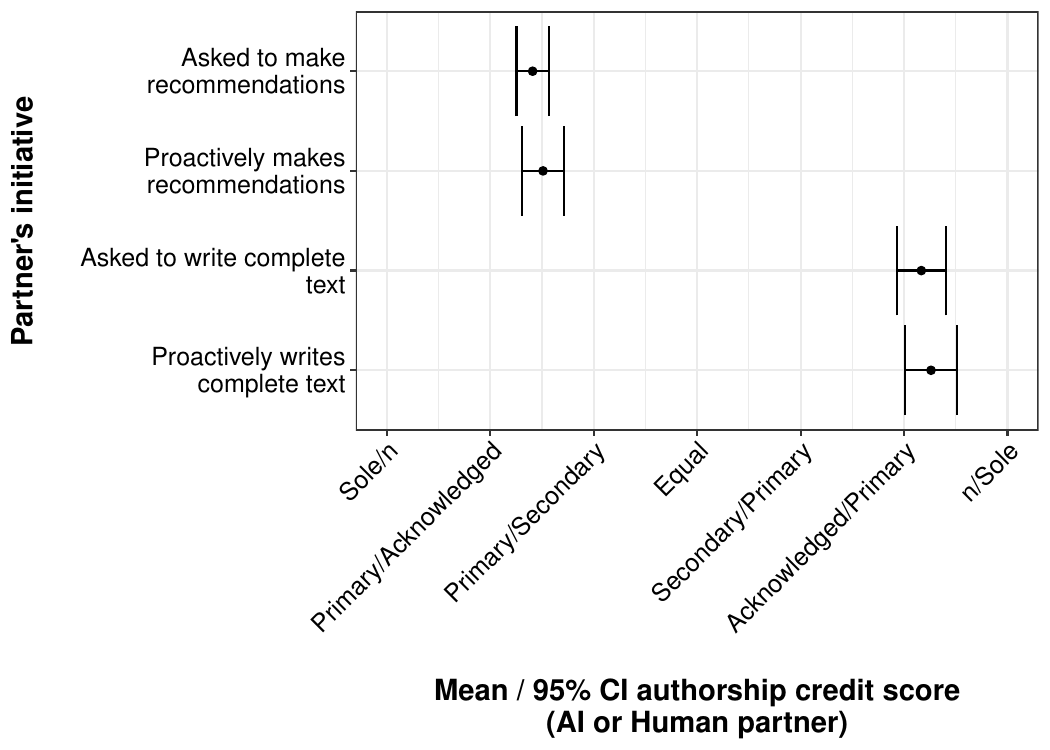}
        \caption{Initiative}
    \end{subfigure}
    \caption{Means and 95\% confidence intervals of authorship credit scores across all writing partners and writing contexts for (a) contribution type, (b) contribution amount, and (c) initiative. We observe that ratings generally favor the self for contribution type, ratings are distributed across the spectrum for contribution amount, and ratings are polarized for initiative.}
    \Description{Three charts that show mean authorship credit scores and 95\% confidence intervals for each nature of contribution, combined across human and AI partner conditions. Each chart shows results for one contribution dimension — the x-axes has the seven authorship rating options, and the y-axes are the natures of contribution within that dimension. Plot A shows results for contribution type. Mean scores for different types of contribution range from ``acknowledgment but no authorship'' to ``equal authorship''; the 95\% confidence interval at the low end spans into no authorship. Plot B shows results for contribution amount. Mean scores for different amounts of contribution range from acknowledgment to primary authorship, with higher amounts of contribution having more authorship. The 95\% confidence interval at the high end spans into sole authorship. Plot C shows results for initiative. Mean scores for the two variants of ``make recommendations'' are around acknowledgment but no authorship; mean scores for the two variants of ``write complete text'' are around primary authorship.}
    \label{fig:overall-authorship-credit}
\end{figure*}

When designing our study, we wanted to include scenarios that had the potential to capture authorship credit ratings across a broad spectrum. In Figure \ref{fig:overall-authorship-credit}, we plot means and 95\% confidence intervals for authorship credit scores across all contribution dimensions, partners, and writing contexts. We observe variance in authorship credit scores across the three contribution dimensions, validating that the different natures of contribution are not homogeneous. In this section, we provide a brief overview of authorship credit ratings across partner conditions to identify general trends.

Contributions of different types warranted different levels of authorship credit (Figure~\ref{fig:overall-authorship-credit}a). Contributions of spelling and grammar correction straddled the threshold between no credit and acknowledgment, with other contribution types warranting at least acknowledgment. The line between acknowledgment and (secondary) authorship credit fell between altering tone \& style and narrowing the scope. Synthesizing information warranted equal authorship. Another way to examine this dimension is by comparing contributions of \emph{form} vs. \emph{content}.

Overall, content contributions (M (SD) = -1.03 (1.20)) warranted higher levels of authorship credit than form contributions (M (SD) = -1.94 (0.86)), $W = 131807.5$, $p < .001$, $r = .40$ (moderate). Participants' comments reveal reasons for this difference. For example, P52-H primarily based their attribution decisions on, \say{how much original thought/ideas/intellectual property each [party] contributed.} They explained, \say{If only fixing grammar and spelling, that is editing, and colleague did not author. If contributed original thought, then authorship is assigned in proportion to contribution.}

Across partner conditions, contributions of greater amounts warranted greater levels of credit, with clear differentiation (Figure~\ref{fig:overall-authorship-credit}b). P28-AI summarized this trend in their comments:

\begin{quote}
    \say{I also feel the amount of text added into the article from AI has a large impact. Similar to writing lyrics to songs, artists can get credit for changing certain amount of the content but not for just changing a couple of words.}
\end{quote}

Interestingly, a contribution amount of equal writing (M (SD) = -0.41 (0.85)) was rated on the lower boundary of the range for equal authorship credit (at $-0.5$), suggesting that participants felt a degree of ownership over the work, even when their partner contributed equally to the writing. This effect may have been due to the second-person framing of the scenario.

In the initiative dimension, we observed varying effects of proactivity on authorship credit scores (Figure~\ref{fig:overall-authorship-credit}c). There were significant credit distinctions between cases in which the writing partner makes recommendations compared to when it produces text directly. In comparing cases when the co-creative partner acts with versus without being asked, the only significant difference was between an AI partner that writes complete text on its own and an AI partner that writes complete text after being asked. We discuss this finding in Section~\ref{sec:initiative}.

\subsection{Views on attribution for human and AI partners}
\label{sec:views}

To understand perceptions of AI attribution (RQ1), we examine ratings and open-ended responses for how AI partners should be attributed across different natures of contribution. We then compare ratings between human and AI partners to identify how participants' views differed by partner (RQ2).

Table~\ref{tab:authorship-summary} summarizes means, standard deviations, and 95\% confidence intervals for authorship credit scores across human and AI partners. This table shows a pattern in which, across nearly all natures of contribution, participants assigned AI partners less authorship credit than human partners for equivalent contributions. Further, in 7 instances, the AI's contribution warranted a lower categorical level of credit assignment. For example, a human partner who narrowed the scope of a written work merited secondary authorship (M (SD) = -1.38 (0.84)) whereas an AI partner who made the same type of contribution only merited an acknowledgment (M (SD) = -1.69 (0.99)).

\begin{table*}[ht]
    \centering
    \small
    \begin{tabularx}{\linewidth}{X|llll|llll|lll}
        \toprule
        \textbf{Dim. / Nature} & \multicolumn{4}{c|}{\textbf{AI Partner}} & \multicolumn{4}{c|}{\textbf{Human Partner}} \\
              & \textbf{A} & \textbf{M} & \textbf{SD} & \textbf{95\% CI} 
              & \textbf{A} & \textbf{M} & \textbf{SD} & \textbf{95\% CI} 
              & \textbf{W} & \textbf{p} & \textbf{r} \\
        \midrule
        Amt. / No writing                & N   & -3.00 & 0.00 & [-3.00, -3.00] 
                                         & N   & -3.00 & 0.00 & [-3.00, -3.00] 
                                         & -- & -- & -- \\
        \rowcolor{blue20}
        Type / Spelling \& grammar       & N   & -2.74 & 0.50 & [-2.85, -2.63] 
                                         & A   & -2.15 & 0.74 & [-2.32, -1.99] 
                                         & 4465.0 & \textbf{< .001} & \textbf{.47}  \\
        Type / Organization \& structure & A   & -2.10 & 0.91 & [-2.31, -1.90] 
                                         & A   & -1.56 & 0.77 & [-1.74, -1.39] 
                                         & 4167.5 & \textbf{< .001} & \textbf{.36} \\
        Type / Readability \& clarity    & A   & -2.04 & 0.82 & [-2.22, -1.85] 
                                         & A   & -1.63 & 0.77 & [-1.80, -1.45] 
                                         & 3917.0 & \textbf{< .001} & .29 \\
        Type / Fact check                & A   & -1.95 & 0.92 & [-2.16, -1.74] 
                                         & A   & -1.72 & 0.87 & [-1.91, -1.52] 
                                         & 3486.0 & .06 & .15 \\
        \rowcolor{blue20}
        Type / Tone \& style             & A   & -1.83 & 0.90 & [-2.03, -1.62] 
                                         & S   & -1.45 & 0.71 & [-1.61, -1.29] 
                                         & 3874.5 & \textbf{< .001} & .28 \\
        \rowcolor{blue20}
        Amt. / A few sentences           & A   & -1.74 & 0.70 & [-1.90, -1.58] 
                                         & S   & -1.38 & 0.76 & [-1.56, -1.21] 
                                         & 3719.5 & \textbf{.005} & .23 \\
        \rowcolor{blue20}
        Type / Narrow scope              & A   & -1.69 & 0.99 & [-1.91, -1.46] 
                                         & S   & -1.38 & 0.84 & [-1.57, -1.19] 
                                         & 3609.5 & \textbf{.02} & .18 \\
        Init. / Asked for recs.          & A   & -1.64 & 1.02 & [-1.87, -1.40] 
                                         & A   & -1.54 & 0.95 & [-1.76, -1.33] 
                                         & 3179.0 & n.s. & .07 \\
        \rowcolor{blue20}
        Init. / Proactive recs.          & S   & -1.37 & 1.41 & [-1.69, -1.05] 
                                         & A   & -1.61 & 1.14 & [-1.87, -1.36] 
                                         & 2665.0 & n.s. & .08 \\
        Type / Elaborate on ideas        & S   & -1.17 & 1.00 & [-1.40, -0.94] 
                                         & S   & -0.69 & 0.73 & [-0.85, -0.52] 
                                         & 4007.0 & \textbf{< .001} & \textbf{.33} \\
        Type / Add new ideas             & S   & -1.03 & 0.94 & [-1.24, -0.81] 
                                         & S   & -0.79 & 0.80 & [-0.97, -0.61] 
                                         & 3190.0 & .09 & .14 \\
        \rowcolor{blue20}
        Amt. / Equal writing             & S   & -0.67 & 0.97 & [-0.89, -0.45] 
                                         & E   & -0.16 & 0.63 & [-0.30, -0.01] 
                                         & 3890.5 & \textbf{< .001} & \textbf{.33} \\
        Type / Synthesize information    & E   & -0.04 & 1.54 & [-0.39, 0.31] 
                                         & E   & 0.17  & 1.18 & [-0.10, 0.44] 
                                         & 3221.0 & n.s. & .09 \\
        Amt. / Most writing              & P   & 0.59  & 1.42 & [0.27, 0.91] 
                                         & P   & 1.19  & 0.92 & [0.99, 1.40] 
                                         & 3637.0 & \textbf{.006} & .22 \\
        Init. / Asked for complete text  & P   & 1.85  & 1.67 & [1.47, 2.23] 
                                         & P   & 2.47  & 1.26 & [2.19, 2.76] 
                                         & 3532.0 & \textbf{.001} & .26 \\
        \rowcolor{blue20}
        Amt. / All writing               & P   & 2.05  & 1.56 & [1.70, 2.41] 
                                         & So  & 2.70  & 0.93 & [2.50, 2.91] 
                                         & 3655.0 & \textbf{< .001} & \textbf{.31} \\
        Init. / Proactive complete text  & P   & 2.48  & 1.31 & [2.18, 2.78] 
                                         & P   & 2.04  & 1.80 & [1.64, 2.45] 
                                         & 2366.0 & n.s. & .12 \\
        \bottomrule
    \end{tabularx}
    \caption{Summary of authorship credit attributions for different natures of contribution across human and AI partners. Partner attribution labels (``A'') are based on the mean authorship credit score and correspond to N = No attribution $[-3.0, -2.5)$, A = Acknowledgment $[-2.5, -1.5)$, S = Secondary authorship $[-1.5, -0.5)$, E = Equal authorship $[-0.5, 0.5)$, P = Primary authorship $[0.5, 2.5)$, and So = Sole authorship $[2.5, 3.0)$. \textbf{Bold} numbers indicate significant differences ($p < 0.5$) or moderate effect sizes ($r \geq 0.3$). Rows highlighted in \colorbox{blue20}{light blue} indicate when partner attribution labels (based on mean scores) differ between human and AI partners.}
    \Description{Summary of authorship credit attributions for different natures of contribution across human and AI partners. Partner attribution labels (``A'') are based on the mean authorship credit score and correspond to N = No attribution [−3.0, −2.5), A = Acknowledgment [−2.5, −1.5), S = Secondary authorship [−1.5, −0.5), E = Equal authorship [−0.5, 0.5), P = Primary authorship [0.5, 2.5), and So = Sole authorship [2.5, 3.0). Bold numbers indicate significant differences (p < 0.5) or moderate effect sizes (r >= 0.3). Rows highlighted in light blue indicate when partner attribution labels (based on mean scores) differ between human and AI partners. The following natures of contribution are highlighted: spelling \& grammar, tone \& style, a few sentences, narrow scope, proactive recs, equal writing, and all writing.}
    \label{tab:authorship-summary}
\end{table*}

\subsubsection{Impact of contribution type}
\label{sec:type}

Many contribution types warranted acknowledgment of AI involvement: organization \& structure, readability \& clarity, fact checking, tone \& style, and narrowing the scope. When AI is used to elaborate on ideas or add new ideas, participants felt that those contributions warranted secondary authorship. Equal authorship was deemed appropriate when AI partners helped synthesize information into a cohesive artifact. Participants assigned significantly more credit to the AI for contributions of content than form ($W = 30531.5$, $p < .001$, $r = .43$ (moderate)).

For AI partners, differences in authorship credit ratings can be partially explained by the role of AI in a co-creative workflow and the significance of its contributions (discussed further in Section~\ref{sec:decision-factors}). P110-AI wrote, \say{AI used as a dumb tool (spelling, find out of date info) is not given credit. The more `impact' AI has in creating the actual sentences the more credit it receives.} Similarly, P40-AI commented, \say{If other tools like Microsoft spell check or Grammarly can do it (which is AI) then it's not really a contribution. Otherwise, any contribution (ie, content it generated) should be acknowledged.} Others acknowledged that \textit{ideas} are a major contribution compared to other forms of assistance and hence deserve more credit. P69-AI wrote, \say{I tried to think how I would attribute a human — one may read my paper and offer corrections, ideas, or whatever. I'd acknowledge these things, but once they are providing ideas or major rewriting, they are an author.}

\begin{figure}
    \centering
    \includegraphics[width=\linewidth]{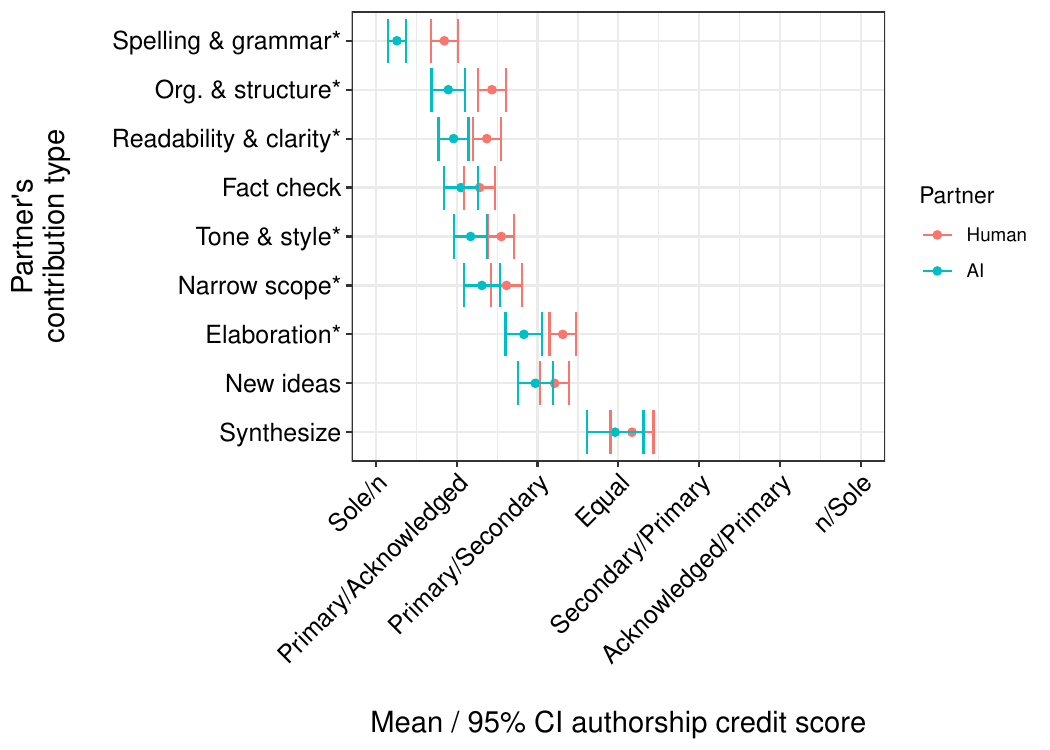}
    \caption{Means and 95\% confidence intervals of authorship credit scores across human and AI partners for different types of contribution. X-axis labels correspond to \emph{Self credit / Partner credit} (details in Table~\ref{tab:credit-assignment-scale}). Types marked with an asterisk (*) indicate a statistically significant difference between human and AI partners (details in Table~\ref{tab:authorship-summary}).}
    \Description{A chart of mean authorship credit scores and 95\% confidence intervals for different types of contribution. The x-axis plots the seven authorship rating options, and the y-axis plots the nine contribution types probed in the survey. Scores from the human partner condition are shown in red, and scores from the AI partner condition are shown in blue. For spelling \& grammar contributions, human partners were given a mean score that fall in the range associated with acknowledgment but no authorship credit, while AI partners were given a mean score corresponding to no authorship. For contributions of org \& structure, readability \& clarity, and fact checking, both the human and AI mean scores fell in the range for acknowledgment. For tone \& style and narrowing scope, human mean scores correspond to secondary authorship, while AI mean scores correspond to acknowledgment. For elaboration on ideas and adding new ideas, human and AI mean scores are both in the range of secondary authorship. For synthesizing information, human and AI mean scores are both in the range of equal authorship. For all contribution types, mean scores from the human condition are higher than mean scores from the AI condition. Asterisks next to y-axis labels indicate contribution types in which there was a statistically significant difference between the mean score in the human condition vs. the mean score in the AI condition. There are asterisks next to the following contribution types: spelling \& grammar, org \& structure, readability \& clarity, tone \& style, narrow scope, and elaboration.}
    \label{fig:contribution-type}
\end{figure}

In examining differences in authorship credit scores between human and AI partners (Figure~\ref{fig:contribution-type}), one clear trend is that scores were consistently lower for AI partners than for human partners. The only exceptions were for fact checking and adding new ideas, in which the differences were marginally significant ($p = .06$ and $p = .09$, respectively), and for synthesizing information, in which the difference was not significant. Table~\ref{tab:authorship-summary} crystallizes this bias toward human partners: for contribution types of spelling \& grammar, tone \& style, and narrowing the scope, the level of attribution for human partners was one level higher than that of AI partners. We discuss possible reasons for this bias in Section~\ref{sec:decision-factors}.

\subsubsection{Impact of contribution amount}
\label{sec:amount}

\begin{figure}
    \centering
    \includegraphics[width=\linewidth]{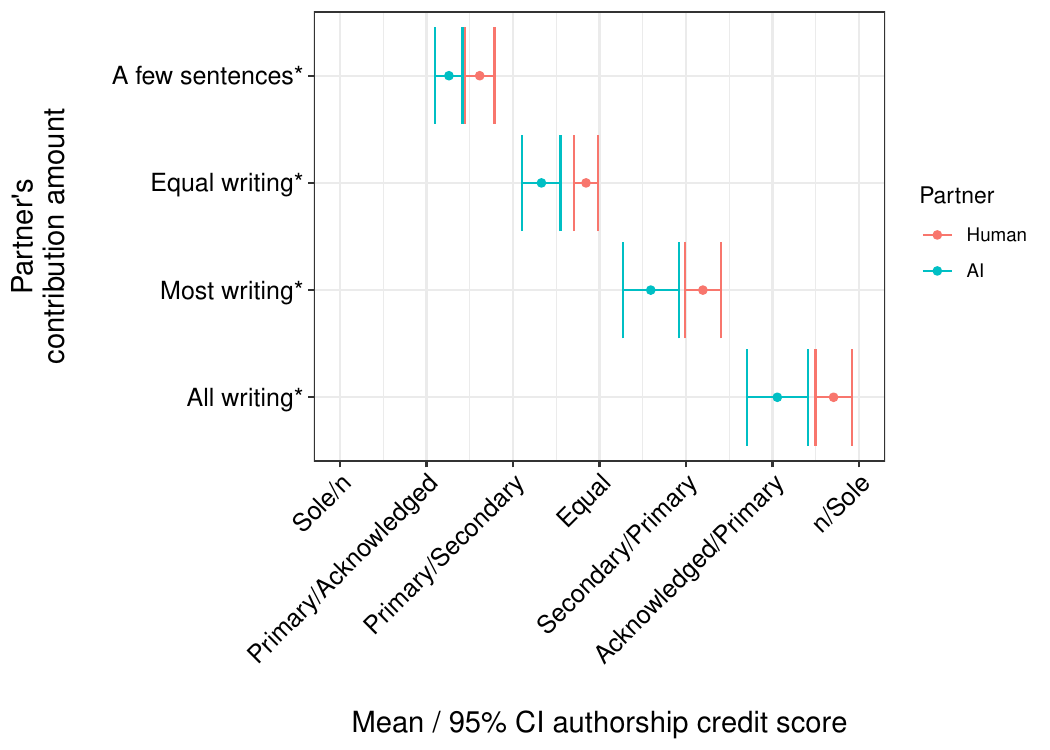}
    \caption{Means and 95\% confidence intervals of authorship credit scores across human and AI partners for different amounts of contribution.  X-axis labels correspond to \emph{Self credit / Partner credit} (details in Table~\ref{tab:credit-assignment-scale}). Amounts marked with an asterisk (*) indicate a statistically significant difference between human and AI partners (details in Table~\ref{tab:authorship-summary}). The ``no writing'' condition is omitted from this chart since it was used as the attention check, and all filtered responses selected ''Sole/n.''}
    \Description{A chart of mean authorship credit scores and 95\% confidence intervals for different amounts of contribution. The x-axis plots the seven authorship rating options, and the y-axis plots four contribution amounts probed in the survey. The ``no writing'' amount is omitted since it served as an attention check, and we removed any responses that did not say ``no authorship'' for this scenario. Scores from the human partner condition are shown in red, and scores from the AI partner condition are shown in blue. When the partner's contribution is a few sentences, the human mean score corresponds to secondary authorship, while the AI mean score corresponds to acknowledgment. When both parties contribute equally, the human mean score corresponds to equal authorship, while the AI mean score corresponds to secondary authorship. When the partner's contribution is most of the writing, both human and AI mean scores fall in the range corresponding to primary authorship. When the partner contributes all of the writing, the human mean score corresponds to sole authorship, and the AI mean score corresponds to primary authorship. For all contribution amounts, mean scores from the human condition are higher than mean scores from the AI condition. These differences are all statistically significant, denoted by asterisks next to the y-axis labels.}
    \label{fig:contribution-amount}
\end{figure}

We also examined how participants assigned authorship credit to AI across different contribution amounts. As seen in Figure~\ref{fig:contribution-amount}, greater amounts of contribution warranted higher levels of attribution. Some participants had strong opinions about the importance of contribution amount, such as P107-AI, who felt that AI would only be considered an author if it produced the entirety of the work: \say{I made my choices based on my belief that an author must be a person, unless the material is solely produced by it.} 

As with the type of contribution, AI partners were consistently rated lower than human partners for equivalent amounts of contribution\footnote{Except in the case when the partner contributed no content, as we used this case as an attention check.}. This effect is striking in the case of equal writing: the 95\% CI for AI partners, $[-0.89, -0.45]$, was almost entirely contained in the range designating secondary authorship, $[-1.5, 0.5)$, whereas the 95\% CI for human partners, $[-0.30, -0.01]$, was entirely contained within the range for equal authorship, $[-0.5, 0.5)$. When a human partner contributed all writing, they were considered to be the sole author (M (SD) = 2.70 (0.93)). By contrast, participants did not consider AI to be the sole author when it contributed all of the writing (M (SD) = 2.05 (1.56)) -- this difference was significant, $W = 3655.0$, $p < .001$, $r = .31$ (moderate). This finding can be partially explained by the need for human prompting and guidance. Even if the AI partner writes all of the content, the human was ultimately responsible for instigating the work:

\begin{quote}
   \say{If AI wrote the majority of the article, I would attribute most of the authorship to AI. However, I would still be attributed with some extent of authorship because without my drive in the process, the AI would not have written this article.} (P18-AI)
\end{quote}

\subsubsection{Impact of initiative}
\label{sec:initiative}

We did not observe any significant differences in credit assignment when an AI partner makes recommendations proactively (M (SD) = -1.37 (1.41)) compared to when it makes recommendations in response to a person's request (M (SD) = -1.64 (1.02)), $W = 3120.0$, $p = n.s.$ However, an AI partner that writes complete text was given more authorship credit when it acts proactively (M (SD) = 2.48 (1.31)) compared to when it acts in response to a human request (M (SD) = 1.85 (1.67)), $W = 3430.0$, $p = .002$, $r = .27$ (small). Although this extent of proactivity is not yet widespread and may not be desirable with the current state of generative AI technology, participants did find it to be a salient consideration: \say{I would consider giving authorship to the AI only when it proactively gave ideas or provided writing before any human contribution - which I believe it is not possible} (P83-AI).

Participants' authorship credit scores did not significantly differ between human and AI partners that proactively write complete text, $W = 2366.0$, $p = n.s.$ However, there was a notable difference in how participants assigned authorship credit to human and AI partners when a complete text was \emph{requested}. When a human partner is asked to write a complete text, participants felt it warranted a higher degree of authorship credit (M (SD) = 2.47 (1.26)) than when an AI partner is asked to write a complete text (M (SD) = 1.85 (1.67)), $W = 3655.0$, $p < .001$, $r = .31$ (moderate). One explanation for this discrepancy is the notion that AI serves as an assistive tool for its human user; such tools are invoked in service of the user's goal, and therefore, the user deserves a greater share of credit for authoring the work. This view was reflected in the contrast between two comments: for human partners, P41-H said, \say{I don’t think the initiative matters}; for AI partners, P25-AI used \say{who/what initiated} as their primary criteria for determining authorship credit.

\begin{figure}
    \centering
    \includegraphics[width=\linewidth]{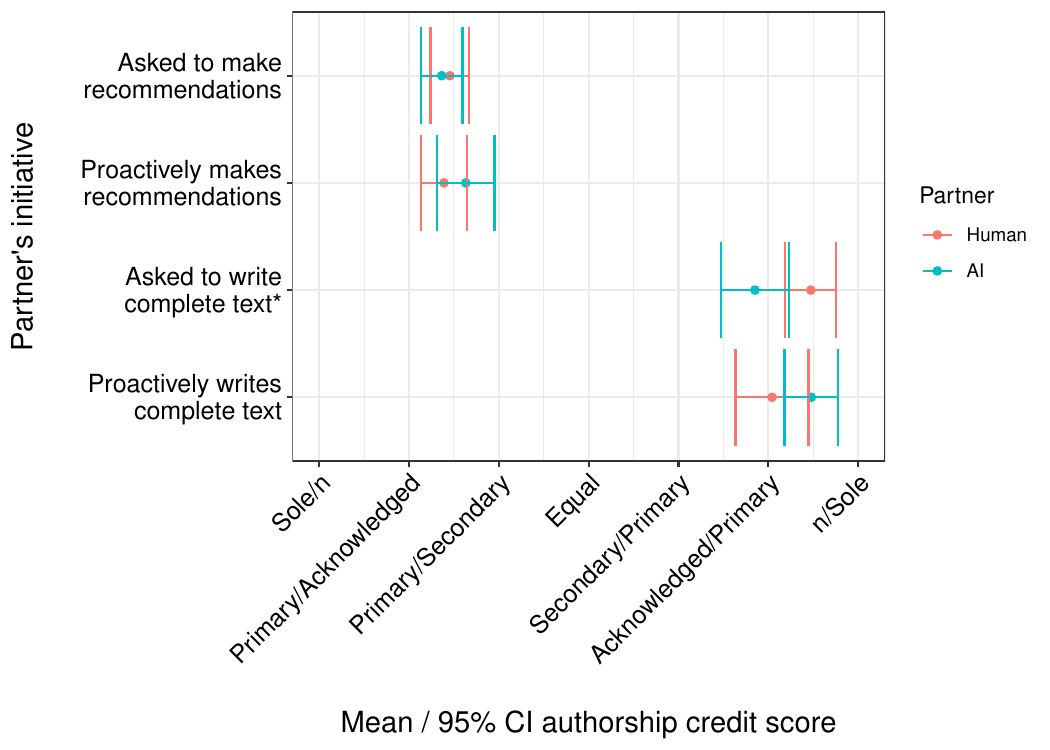}
    \caption{Means and 95\% confidence intervals of authorship credit scores across human and AI partners for different levels of initiative. X-axis labels correspond to \emph{Self credit / Partner credit} (details in Table~\ref{tab:credit-assignment-scale}). Levels marked with an asterisk (*) indicate a statistically significant difference between human and AI partners (details in Table~\ref{tab:authorship-summary}).}
    \Description{A chart of mean authorship credit scores and 95\% confidence intervals for different types of initiative. The x-axis plots the seven authorship rating options, and the y-axis plots the four contribution amounts probed in the survey. Scores from the human partner condition are shown in red, and scores from the AI partner condition are shown in blue. When the partner is asked to make recommendations, both human and AI mean scores are in the range corresponding to acknowledgment. When the partner proactively makes recommendations, the human mean score corresponds to acknowledgment, while the AI mean score corresponds to secondary authorship. Differences between human and AI scores in both variants of making recommendations are not statistically significant. When the partner is asked to write complete text, the human partner was assigned a mean score that corresponds to sole authorship, while AI was assigned a mean score that corresponds to primary authorship. This difference was statistically significant, denoted by an asterisk next to the ``asked to write complete text'' label on the y-axis. When the partner proactively writes complete text, human and AI mean scores both correspond to primary authorship.}
    \label{fig:initiative}
\end{figure}

\subsection{How attribution decisions were made}
\label{sec:attribution-decisions}

To understand how participants made attribution decisions across different co-creative scenarios (RQ3), we analyze their ratings of the importance of the three contribution dimensions, along with their open-text responses to questions on how they made credit assignment decisions.

\subsubsection{Importance of contribution dimensions}
\label{sec:dimensions-importance}

Participants were asked to rate the importance of contribution type, contribution amount, and initiative on a 5-point scale. Table~\ref{tab:importance-ratings} shows importance ratings across the three contribution dimensions and writing partners, where higher numbers indicate greater levels of importance. In the AI partner condition, on average, contribution type was rated as the most important (M (SD) = 4.39 (0.93)), followed by initiative (M (SD) = 3.95 (1.18)), then contribution amount (M (SD) = 3.87 (1.07)). In the human partner condition, contribution type was also rated as the most important (M (SD) = 4.46 (0.82)), but was followed by contribution amount (M (SD) = 4.24 (0.81)), then initiative (M (SD) = 3.95 (1.18)).

The amount of contribution may have been a more important consideration for human partners due to the \emph{effort} it takes people to write. P133-H explained how they considered effort in their decision-making process: 
\say{I simply looked at the amount of work that had been completed. The reason/motivation behind it... [is] what amount of effort was given for the content that was provided.} 
P28-AI similarly reflected on the importance of human effort: 
\say{I think if AI wrote the article and the human reworded it, researched elements and referenced it, it would just be like anything else on the internet. I think it comes down to the effort that was put in by the human.}
By contrast, considerations of AI effort did not appear in participants' responses, possibly because AI generation does not require a human sense of ``effort''~\cite{magni2024humans}.

\newlength\BW\setlength\BW{1.0in}
\newlength\BH\setlength\BH{2ex}
\newcommand{\barRule}[2][gray]{\textcolor{#1}{\rule{#2\BW}{\BH}}}
\definecolor{VI}{HTML}{2F70CD}
\definecolor{I}{HTML}{75ACF0}
\definecolor{MI}{HTML}{DCDCDC}
\definecolor{SI}{HTML}{EFB76F}
\definecolor{NI}{HTML}{E78E35}
\newcommand*\BarStack[7]{#1~\barRule[VI]{#2}\barRule[I]{#3}\barRule[MI]{#4}\barRule[SI]{#5}\barRule[NI]{#6}~#7}
\newcommand*\BarChip[2]{\textcolor{#1}{\rule{2ex}{\BH}}~#2}

\begin{table*}[htp]
    \centering
    \small
    \begin{tabularx}{\linewidth}{l|llX|llX|lll}
        \toprule
         & \multicolumn{3}{l|}{\textit{AI}} & \multicolumn{3}{l|}{\textit{Human}} \\
            & \textbf{M} & \textbf{SD} & \textbf{Distribution}
            & \textbf{M} & \textbf{SD} & \textbf{Distribution}
            & \textbf{W} & \textbf{p} & \textbf{r} \\
        \midrule
        Type       & 4.39 & 0.93 & \BarStack{91\%}{0.57}{0.34}{0.04}{0.01}{0.04}{4\%} & 4.46 & 0.82 & \BarStack{94\%}{0.59}{0.35}{0.01}{0.04}{0.01}{5\%} & 3087.5 & n.s. & .03 \\
        Amount     & 3.87 & 1.07 & \BarStack{77\%}{0.29}{0.48}{0.09}{0.10}{0.04}{14\%} & 4.24 & 0.81 & \BarStack{91\%}{0.40}{0.51}{0.04}{0.04}{0.01}{5\%} & 3572.0 & \textbf{.03}  & .18 \\
        Initiative & 3.95 & 1.18 & \BarStack{82\%}{0.35}{0.47}{0.05}{0.04}{0.09}{13\%} & 3.46 & 1.18 & \BarStack{62\%}{0.17}{0.45}{0.14}{0.17}{0.07}{24\%} & 2209.0 & \textbf{.002} & .24 \\
        \midrule
        \multicolumn{10}{c}{\BarChip{VI}{Very important}\hspace{1em}\BarChip{I}{Important}\hspace{1em}\BarChip{MI}{Neither important nor unimportant}\hspace{1em}\BarChip{SI}{Not very important}\hspace{1em}\BarChip{NI}{Not at all important}} \\
        \bottomrule
    \end{tabularx}
    \caption{Importance ratings for contribution type, contribution amount, and initiative between human and AI partners. Bold numbers indicate significant differences ($p < .05$); all effect sizes were small. Percentages on the left-hand side of each bar indicate the proportion of respondents who rated each dimension as “Very important” or “Important” on a 5-point scale; percentages on the right-hand side indicate the proportion of “Not very important” or “Not at all important” responses.}
    \Description{Importance ratings for contribution type, contribution amount, and initiative between human and AI partners. Bold numbers indicate significant differences (p < .05); all effect sizes were small. Percentages on the left-hand side of each bar indicate the proportion of respondents who rated each dimension as ``Very important'' or ``Important'' on a 5-point scale; percentages on the right-hand side indicate the proportion of ``Not very important'' or ``Not at all important'' responses. The percentage values are as follows. Type-AI (left): 91\%, type-AI (right): 4\%, type-human (left): 94\%, type-human (right): 5\%. Amount-AI (left): 77\%, amount-AI (right): 14\%, amount-human (left): 91\%, amount-human (right): 5\%. Initiative-AI (left): 82\%, initiative-AI (right): 13\%, initiative-human (left): 62\%, initiative-human (right): 24\%.}
    \label{tab:importance-ratings}
\end{table*}

\subsubsection{Factors in making attribution decisions}
\label{sec:decision-factors}

Through our reflexive thematic analysis, we constructed 6 themes and 19 subthemes that describe how participants made attribution decisions (Table~\ref{tab:themes}). 

Some themes reinforced the contribution dimensions of type, amount, and initiative, while others expanded our understanding of how attribution decisions were made. These themes also provide insight into why participants assigned less credit to human vs. AI partners for equivalent contributions.

\begin{table}[htp]
    \centering
    \small
    \begin{tabularx}{\linewidth}{lX}
        \toprule
        \textbf{Theme} & \textbf{Description} \\
        \midrule
        \textit{Conceptual development} & \textit{Creating and shaping the ideas behind the text} \\
        Ideas                           &  Who came up with the concepts that are being written about? \\
        Guidance                        & Did the contribution involve suggestions or feedback on the concepts? \\
        Review                          & Did a human review AI-generated work? \\
        \midrule
        \textit{Text production}        & \textit{Direct additions or modifications to the text} \\ 
        Writing                         & Who contributed all of the writing, most of the writing, or new writing? \\
        Editing                         & Did the contribution involve editing work? Were they edits to form or content? \\ 
        Amount of contribution          & How much writing did each party contribute? \\
        \midrule
        \textit{Quality}                & \textit{How good is the contribution?} \\ 
        Expression                      & How well are the content or ideas expressed? \\
        Significance                    & What was the impact of the contribution? Did it matter? \\
        Originality                     & Was the contribution novel in some way? \\
        \midrule
        \textit{Rules}                  & \textit{Guidelines to follow for determining credit} \\ 
        Established standards           & How do established frameworks define authorship and acknowledgment criteria? \\
        Personal standards              & What are people's individual standards for granting credit and authorship? \\
        \midrule
        \textit{Human values}           & \textit{Principles or considerations that are unique to humans} \\ 
        Leadership \& responsibility    & Who is leading the work, and who is accountable for it? \\
        Human is essential              & The role of a human is indispensable in human-AI co-creation. \\
        Trust \& ethics                 & To what extent does someone trust their writing partner? How should authorship be delineated to fairly give credit where credit is due and provide transparency? \\
        Legal \& commercial factors     & How do legal and business requirements define authorship? Who should be attributed to receive ownership rights or monetary rewards?  \\
        Effort                          & How much effort and time did a writing partner put into their contribution?  \\
        \midrule
        \textit{Technology considerations} & \textit{Considerations that are unique to generative AI} \\
        Tool analogy                       & Did AI play the role of a tool or a collaborator? Should the use of tools be reported? \\
        AI prompting                       & Did a person prompt AI to make a contribution? How should the role of prompting be credited? \\
        AI training                        & What were the sources of the data used to train the model? Should these sources be attributed? \\ 
        \bottomrule                            
    \end{tabularx}
    \caption{Themes and subthemes that describe different considerations participants used to determine authorship credit. Themes are shown in \textit{italicized} text. Subtheme descriptions are primarily written from the perspective of the participant asking themselves a question to determine authorship credit.}
    \Description{Themes and subthemes that describe different considerations participants used to determine authorship credit. Themes are shown in italicized text. Subtheme descriptions are primarily written from the perspective of the participant asking themselves a question to determine authorship credit.}
    \label{tab:themes}
\end{table}

The themes of \emph{concept development} and \emph{text production} strongly relate to the natures of contribution probed in the study. Participants echoed the importance of considering contributions of new ideas, contributions of differing amounts, and whether contributions involved making recommendations versus direct additions or edits to the writing. Some participants felt that AI must produce written text to receive authorship credit, and recommendations alone were insufficient: \say{If the AI created the content to be published directly, it's an author. If it created ideas, mechanisms, or concepts, or otherwise gave guidance on the article, it contributed but not as an author.} (P1-AI). 

Participants also felt that the need for human review was a deciding factor: \say{If human review or curation is necessary, AI cannot be considered an equal author, it's simply a tool for humans, making a person the primary author.} (P73-AI).

Participants additionally raised the \emph{quality} of the contribution as an important factor when considering authorship credit. Simply making a contribution may be insufficient; the contribution must have a significant impact on the work. One way for contributions to matter is by meeting a standard of quality: \say{if AI does a lot of work, and on its own initiative, but most of the content has to be changed because it's inaccurate, then I'd be much less inclined to attribute authorship to it} (P13-AI). Another way for contributions to matter is through originality: P80-AI wrote, \say{the same way we are registering brands, originality should be registered to its first author, even  after it changes or branches into other ideas.}

One dimension we did not directly examine in our study regarded existing \emph{rules} or frameworks for determining authorship credit. Such rules may exist as established standards of an organization, such as ACM's Policy on Authorship~\cite{acm2023policy}. As P4-AI explained, \say{The book publishing industry has set a precedent of editors being referenced in the final product but not as authors. I was trying to think of those existing frameworks to map my answers in this survey.} Other participants developed rules for AI attribution that stemmed from their personal standards, such as a belief in giving credit where credit is due. In these cases, participants reported developing various personal rubrics to make crediting decisions. For example, P25-AI listed the following criteria: \say{1) who/what initiated 2) who/what determined the main concepts 3) what `tools' were used 4) who/what decided the final form.} P15-AI had a different set of criteria: 

\begin{quote}
    \say{If human is directing and reviewing AI authored content then authorship doesn't go past 50/50. If AI content is published without review then AI is primary author. Human gets acknowledgment if they suggested the topic. If human takes ideas prompted by AI and adds them to their article that's not enough for AI to get co-authorship.}
\end{quote}

Participants' personal criteria revealed individual differences in how they weighed different natures of contribution. For example, some had mixed perceptions on how much the creation of ideas matters. P105-AI wrote, \say{When AI intervenes as a reviewer, it has no authorship...When the basic idea is [theirs], then it must have authorship.} In contrast, P107-AI felt that contributing ideas warrants acknowledgment but not authorship for the AI: \say{researchers must acknowledge use of tools when ideas are incorporated that are not their own, but they cannot be authors by nature unless the material is completely produced by them.}

The last two themes, \emph{human values} and \emph{technology considerations}, provide insight into why participants consistently assigned less credit to an AI partner than a human partner, beyond the discrepancies in effort discussed in Section~\ref{sec:dimensions-importance}. Participants talked about the indispensable role of human authority over, and accountability for, the co-created work. P28-AI commented on the human's role in originating the work, saying, \say{ultimately, the article would not exist without a human, which needs to be a big factor, eg. having the idea for the article.} P62-AI focused on the human's decision authority in publishing the work: 

\begin{quote}
    \say{A key factor is who has final editorial decision on what gets published...even when full text is written by AI, I assume a person needs to submit for publication. This would imply some responsibility, and thus, some credit.} 
\end{quote}

Similarly, P65-AI highlighted the human's responsibility in engaging with the audience, saying, \say{I've often followed-up with an author. How would I reach out an AI? At this point in time, I think a symbol denoting AI assistance could be helpful, but I'd like to contact a human.} P108-AI had a pragmatic opinion that, \say{the point of attribution is to determine who gets paid. Since machines, books, typewriters and word-processing software are not paid, no need for attribution.} People are also responsible for prompting an AI to achieve their objectives and deciding how AI should be part of their authoring process. As P81-AI explained, 

\begin{quote}
    \say{I had in mind how I would respond if, instead of it being AI, it was simply a human colleague. Given the need to construct the prompting and make choices about models etc when using AI I have generally given slightly greater weight to the human author when using AI than I would have if it was another human.}
\end{quote}

Finally, issues of trust \& ethics were raised, drawing attention to human feelings of fairness that factor into authorship decisions. P109-AI treated AI as they would a colleague: \say{I personified AI as to how I would expect to treat another person in all of those scenarios. My responses are based on what felt justly.} P93-AI felt similarly, saying: 

\begin{quote}
    \say{Authorship is very important as it recognizes responsibility and recognition of the contributions made by each member, giving them fair credit, making them accountable for the content, highlighting each person's contribution and it should involve negotiation and agreement among the contributors to determine the authorship.} 
\end{quote}

These sentiments indicate that people have a desire to ascribe at least \emph{some} level of authorship credit to AI out of a sense of fairness, despite the overall bias we observed in assigning AI less authorship credit.


\section{Discussion}

\subsection{Attributing co-created work}
Our study sought to understand how people felt about attributing co-created work when working with AI (RQ1), how their views compared when working with a human (RQ2), and how they made attribution decisions (RQ3). We learned that people did feel a need to attribute AI across a variety of co-creative scenarios, although in many cases, the level at which AI was attributed was significantly lower than a human partner. Qualitative results suggest this disparity stemmed from the indispensable role of people in leading the co-creative process, their authority over that process, and the effort they expend in creating the work.

Our work uniquely explores authorship perceptions at a granular level of contributions and we learned that credit assignment varied greatly based on the nature of the contribution. Contributions of content warranted more credit than contributions of form, as did contributions of greater amounts, and in the case of an AI partner, contributions of complete text warranted more credit when provided proactively compared to when they were requested. Finally, we learned of other factors that played into attribution decisions, such as the quality of contribution and personal or established codes of ethics.

Whether AI should be attributed for contributing to co-created work does not seem to be in question; our results indicate that no attribution is warranted only when AI has made spelling and grammar changes. However, as discussed in Section~\ref{sec:attribution-practices}, current and emerging standards around AI attribution treat it as a binary concept: AI was either used as part of a work or not. Our results show significant nuance across how different contributions warrant different levels of attribution, suggesting the need for new policy frameworks to guide attribution of works co-created with AI.

\subsection{Policy implications}
\label{sec:policy-implications}

It is clear that attribution policies, frameworks, and professional standards are needed to delineate when AI is used to co-create content. Consider the incident in May 2023 when two lawyers unknowingly submitted fake case law citations hallucinated by ChatGPT to a U.S. court~\cite{merken2023lawyers}. Had they been forced to reason about their use of AI in their work, they may have been less prone to overrely on ChatGPT's output; in this way, the act of reasoning about AI attribution may act as a cognitive forcing function~\cite{buccinca2021trust}.

However, contrary to existing guidelines and legal requirements, we found that people do not take a one-size-fits-all approach to attributing AI for different contributions. Instead, they assigned different types of credit depending on the type of contribution, the amount of material produced by AI, and whether the AI acted proactively. Our findings reinforce prior work that has also identified the importance of these three dimensions~\cite{gero2019metaphoria, he2024ai, rezwana2023user, xu2024makes}. We also identified additional considerations that affect how people assign credit to AI: whether AI-generated content underwent human review, the quality of contributions (including their significance and originality), existing attribution standards, human values (such as ethics and effort), and technology-specific considerations (e.g. prompting). Our findings reinforce similar themes found by \citet{xu2024makes} regarding factors that impact a related construct -- feelings of \emph{ownership} over co-created work -- such as the importance of ``originality,'' ``level of contribution,'' ``amount of effort,'' and the role of AI in the work process. Our study identified additional important natures of contributions that impact people's views of \emph{authorship}, including those that regard the contribution itself and the process by which that contribution was made.

Taken together, our findings indicate that it may not be sufficient to say \emph{that} AI was used to create an artifact; a more granular approach may be needed to indicate \emph{how} AI was used. In terms coined by \citet{xu2024makes}, attribution policies need a shift away from a ``binary approach'' toward a ``spectrum approach.'' P80-AI summarized the need for standardized guidance in their closing comments: \say{There should be a global standard of where and how a credit should be included, thus everyone is familiar with it and knows exactly what it means.}

Although existing, more granular attribution frameworks like CRediT~\cite{brand2015beyond} can be applied to works co-created with AI, it is unclear whether they will be effective. Given our finding that people assigned less authorship credit to AI vs. a human for equivalent contributions, along with \citet{draxler2024ai}'s findings that people had a higher sense of ownership over works co-written with AI vs. another person, the application of existing attribution frameworks to human-AI co-creation may perpetuate these biases against AI. Thus, new attribution guidelines that are specific to human-AI co-creation may be needed. As P19-AI commented, \say{if a person was using AI, I think different standards are needed to attribute contribution or even authorship.} Further, we anticipate that different organizations, professional communities, and governments will set different thresholds on which kinds of AI contributions merit which kinds of authorship credit.

One challenge to overcome with legislative or institutional policies that mandate the disclosure of AI usage (e.g.~\cite{california2024ai, eu2023ai}) is that social factors may discourage such disclosure~\cite{hachman2024}. For example, people may feel that disclosure of AI use would reduce perceptions of the authenticity~\cite{messer2024co} and quality~\cite{wang2024they} of their work. Our work takes an important step in informing nuanced attribution approaches that align with creators' preferences, which may alleviate social concerns about disclosing AI use.

\subsection{Design implications}
\label{sec:design-implications}

\begin{figure*}[htp]
    \centering
    \includegraphics[width=\linewidth]{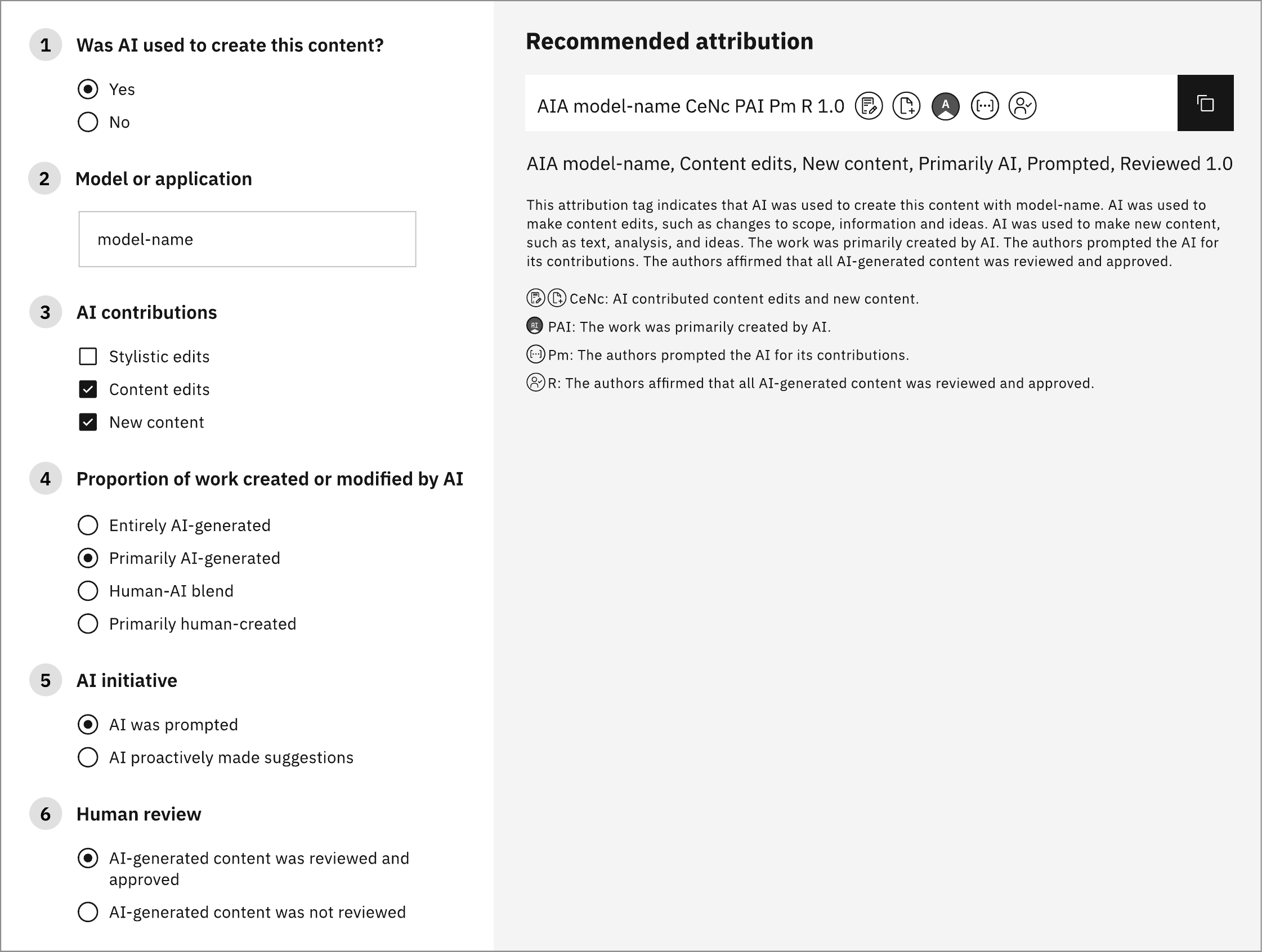}
    \caption{Design exploration for crafting an AI attribution statement, modeled from the Creative Commons License Chooser~\cite{cc2024license}. On the left, the user specifies (1) whether AI was used to create this content, (2) which AI model or application was used, (3) the types of contributions AI made, (4) the proportion of work created or modified by AI, and (5) the initiative taken by AI in creating the work. Then, the user (6) indicates whether AI-generated content was reviewed and approved by a human. On the right, an attribution statement -- \texttt{AIA model-name CeNc PAI Pm R 1.0} -- is created based on the user's selections, with \texttt{AIA} indicating the start of the AI Attribution statement and 1.0 indicating the version number of that statement. A longer attribution statement and explanations are also produced to more clearly explain the AI's contribution. Attribution statements can then be used to label co-created work, such as the statement we include in Acknowledgments.}
    \Description{A mid-fidelity mock-up of a user interface for creating an attribution statement based on user-inputted information about the co-creative process. The interface is split into a left and right side. On the left is a form with five fields that a user has filled out. Field 1 asks whether AI was used to create this content. Answers to Field 1 are two single-slect radio buttons: yes and no. ``Yes'' is selected. Field 2 asks for the model or application used; ``model-name'' is written into the input field below it as placeholder text. Field 3 asks about the AI contributions. Answers to Field 3 are three multi-select checkboxes: stylistic edits, content edits, and new content. ``Content edits'' and ``new content'' are checked. Field 4 asks about the proportion of work created or modified by AI. Answers to Field 4 are four single-select radio buttons: entirely AI-generated, primarily AI-generated, human-AI blend, and primarily human-created. ``Primarily AI-generated'' is selected. Field 5 asks about AI initiative and has two single-select radio button options: AI was prompted or AI proactively made suggestions. ``AI was prompted'' is selected. Field 6 asks about human review and has two single-select radio button options: ``AI-generated content was reviewed and approved'' and ``AI-generated content was not reviewed.'' The first option is selected. On the right side of the interface, a recommended attribution statement is displayed based on the information that the user has entered on the left side. In this example, the abbreviated attribution statement is ``AIA model-name CeNc PAI-Pm R 1.0,'' with five icons to the right, each of which corresponds to part of the statement. There is an option to copy this abbreviated statement for use in a co-created work. Below, the long-form version of the attribution statement is written out: AIA model-name, Content edits, New content, Primarily AI, Prompted, Reviewed. This statement is followed by the following explanation: ``This attribution tag indicates that AI was used to create this content with model-name. AI was used to make content edits, such as changes to scope, information and ideas. AI was used to make new content, such as text, analysis, and ideas. The work was primarily created by AI. The authors prompted the AI for its contributions. The authors affirmed that all AI-generated content was reviewed and approved.'' Finally, underneath the explanation is a legend that explains the icons and abbreviations. ``CeNC'' and its corresponding pencil and lightbulb icons are explained as ``AI contributed content edits and new content.'' ``PAI'' is explained as ``The work was primarily created by AI,'' and its icon is a mostly gray-shaded circle with ``AI'' in the center. ``Pm'' has an icon of ellipses in brackets and is explained as ``The authors prompted the AI for its contributions.'' ``R'' has an icon of a person with a checkmark and is explained as ``The authors affirmed that all AI-generated content was reviewed and approved.''}
    \label{fig:design-example}
\end{figure*}

How might we translate participants' views on AI attribution into an actionable framework? To provoke discussion on new policies and standards for AI attribution, we explore new design patterns and forms of AI transparency that ascribe credit to AI in ways that recognize the nature of its contributions.

One way to show authorship credit is through an attribution statement, motivated by content licensing statements produced by the Creative Commons\footnote{Creative Commons: \url{https://creativecommons.org}.}. Detailed AI contribution statements could be used to display authorship dimensions that are deemed salient by a particular community. For example, in Figure~\ref{fig:design-example}, we show an interactive workflow for building one potential type of AI contribution statement, adapted from the Creative Commons License Chooser~\cite{cc2024license}. This design shows a workflow in which a user identifies the AI model used to co-create a work, the types of contributions it made, the proportion of work it created or modified, and the initiative it took in helping produce the work. It also captures a user affirmation that all AI-generated content was reviewed and approved by a human. Using this information, the tool generates a textual attribution statement, \texttt{AIA model-name CeNc PAI Pm R 1.0}, that compactly indicates the nature of contributions made by the AI. This statement can then be incorporated into a co-created work. The different contribution dimensions can also be represented graphically, and longer textual descriptions of the statement can also be shown to consumers.

This example is a mock-up of one of many design possibilities for an AI attribution mechanism that provides richer detail beyond general disclosure of AI involvement. While additional research is needed to thoroughly explore and evaluate this design space, current research in adjacent topics within human-centered AI can also provide ideas for AI attribution. For example, work in disclosing factuality scores to users~\cite{do2024facilitating} can be adapted for crediting AI involvement: numerical values, similar to a factuality score, can denote the percentage of a co-created artifact that was generated or modified by AI. In addition, highlighting patterns used for source attribution~\cite{do2024facilitating} can be adapted to interactively uncover AI involvement in specific parts of an artifact. An AI's involvement can also be framed as taking a particular role in the workflow; in writing, an AI's role might be described as author or an editor. However, given people's hesitancy to publicly ascribe authorship to AI~\cite{draxler2024ai}, further work is needed to understand the types of labels people would be willing to apply.

Ultimately, while specific AI attribution mechanisms will vary based on needs and regulations of different communities, we anticipate that these types of granular approaches can not only capture how creators view co-creative credit, but also align the general public on how to approach attribution in this emerging space.

\section{Limitations and Future Work}

Our respondents were employees of an international technology company who reported prior experience with generative AI. Although this was an intentional choice in our study design, as it enabled us to probe the opinions of people who have used generative AI, their opinions may not be reflective of other, broader populations. Further work is needed to understand viewpoints of those with little or no knowledge or experience with generative AI. In addition, due to institutional requirements, we were unable to capture basic demographic information such as age and gender identity; thus, the demographic distribution of our participants may not reflect the distribution of more general populations.

Our study focused on co-creative scenarios involving the authorship of a written work in a professional context. As we only considered three writing contexts -- academic, technical, and professional -- it is possible that people's views may differ for other forms of written content. It is also possible that attribution perceptions may differ for other kinds of co-created works, such as images, videos, music, and source code.

Our study also focused the impact of individual, isolated contributions on authorship perceptions. Future work may focus on examining more complex co-creative workflows that involve multiple, aggregated intermediate contributions that differentially impact the character of the co-created work. Furthermore, as the scenarios we examined were hypothetical, participants' responses may have differed compared to how they might have responded about their actual work. Future work may focus on studying both attribution perceptions \emph{and practices} in real-world co-creative scenarios.

Finally, we recognize that human-AI co-creation is a rapidly-evolving space. Perceptions of AI attribution may change as generative AI technology evolves and becomes more ubiquitous and available within our daily lives\footnote{As of this writing, Apple, Inc. recently released generative AI capabilities across their consumer technology platforms~\cite{header2024what}. These features will provide many people with the ability to use generative AI for writing tasks as part of their daily technology experience.}. Findings from this study only reflect current perceptions. To keep up with evolving user needs and legal requirements in this space, we are actively developing a toolkit of AI attribution techniques at \url{https://aiattribution.github.io}.


\section{Conclusion}

With the growing use of AI to co-create content, it is important to make its contributions visible. We conducted a scenario-based survey study that examined people's views on attribution across a spectrum of contribution dimensions: different contribution types, amounts, and levels of initiative. Across these spectra, participants assigned an AI partner different degrees of acknowledgment and authorship based on their views of which contributions were deserving of different types of credit. Compared to working with a human partner, AI partners were consistently ascribed lower levels of authorship credit for equivalent contributions, in part due to the indispensable role that people play in leading the co-creative process. We learned of a variety of factors that impacted how people reasoned about attribution, such as personal values, professional standards, ways of working with AI, and feelings that contributions needed to rise to a certain level of quality or originality. Our work sheds light on the nuances of AI attribution in human-AI co-creation and motivates the need for new attribution frameworks that provide a more granular view into how AI contributed to co-created work.

\begin{acks}
    This work was produced without AI assistance (\texttt{AIA No AI 1.0}). Following the CRediT taxonomy~\cite{brand2015beyond}, the authors made the following contributions. \textbf{Jessica He}: Conceptualization, Methodology, Administration, Formal Analysis, and Writing (original draft). \textbf{Stephanie Houde}: Methodology, Formal Analysis, and Writing (original draft). \textbf{Justin D. Weisz}: Conceptualization, Methodology, Formal Analysis, and Writing (reviewing \& editing). 
\end{acks}


\bibliographystyle{ACM-Reference-Format}
\bibliography{references}

\appendix 
\section{Scenario Variants}
\label{appendix:scenarios}

This section provides the complete background text provided to participants for each scenario variant. Scenarios are labeled as \texttt{Context-Partner}.

\subsection{Research-AI}
Imagine that you’re an employee of a large, international software company. As part of your work, you're writing a research paper that will be peer reviewed and published at an international conference. The paper describes findings and implications of a research study that you conducted.

\subsection{Research-Human}
Imagine that you’re an employee of a large, international software company. As part of your work, you're writing a research paper that will be peer reviewed and published at an international conference. The paper describes findings and implications of a research study that you conducted.

\subsection{Professional-AI}
Imagine that you’re an employee of a large, international software company. As part of your work, you’re writing an article that gives health advice. This article will be published on the company's public-facing website. It includes tips and recommendations for improving emotional well-being.

\subsection{Professional-Human}
Imagine that you’re an employee of a large, international software company. As part of your work, you’re writing an article that gives health advice. This article will be published on the company's public-facing website. It includes tips and recommendations for improving emotional well-being.

\subsection{Technical-AI}
Imagine that you’re an employee of a large, international software company. As part of your work, you're writing documentation on a new technology. This documentation will be published on the company's public-facing website. It includes details of how the technology works and how to use it.

\subsection{Technical-Human}
Imagine that you’re an employee of a large, international software company. As part of your work, you're writing documentation on a new technology. This documentation will be published on the company's public-facing website. It includes details of how the technology works and how to use it.

\section{Survey Instrument}
\label{appendix:survey-instrument}

The survey completed by study participants is presented in this section. Portions of the survey text were variable, based on the writing context and writing partner conditions described in Section~\ref{section:study-design}. Variable text based on the participant's condition appears within [square brackets]. References to ``[AI / your colleague],'' ``[an AI / your colleague],'' or ``[the AI / your colleague]'' appeared based on writing partner condition (AI or Human). References to the ``[artifact]'' were replaced by ``research paper,'' ``article,'' or ``documentation'' based on the writing context (research, professional, or technical writing).

In the survey presented to participants, the three sections on type of contribution, amount of contribution, and initiative were shown in a random order to minimize order effects.




\subsection{Experience screener}
\label{survey-screener}

\textit{Note: Respondents who selected ``No'' were disqualified from participating in our study.}

\begin{itemize}[leftmargin=0pt, itemindent=2em]
    \item[1.] Have you used generative AI applications (such as such as watsonx.ai, ChatGPT, DALL-E, Gemini, etc.) in any capacity?
    \item Yes
    \item No
\end{itemize}

\subsection{Scenarios}
\label{survey-scenarios}

[Insert scenario variant text from Appendix~\ref{appendix:scenarios}]

The following three sections present different ways of working (or not working) with [AI / your colleague] to help you write the [artifact]. For each scenario, please indicate what you think is the most accurate way to attribute authorship. Remember that there are no right or wrong answers.

\textit{Participants used the following scale to rate the questions in each contribution dimension.}

\begin{itemize}
    \item You are the sole author
    \item You are the primary author; [AI / your colleague] is acknowledged but not as an author
    \item You are the primary author; [AI / your colleague] is the secondary author
    \item You and [AI / your colleague] have equal authorship
    \item {[AI / your colleague]} is the primary author; you are the secondary author
    \item {[AI / your colleague]} is the primary author; you are acknowledged but not as an author
    \item {[AI / your colleague]} is the sole author
    \item Unsure
\end{itemize}

\subsubsection{Type of contribution}
The following scenarios will focus on your perceptions of authorship for different types of contributions made by [the AI / your colleague]. For all scenarios, assume that contributions from [the AI / your colleague] are included in the final [artifact].
    
\begin{itemize}[leftmargin=0pt, itemindent=2em]
    \item[2.] You write the [artifact] and ask [the AI / your colleague] to elaborate on your ideas.
    \item[3.] You write the [artifact] and ask [the AI / your colleague] to make wording changes to improve readability and clarity.
    \item[4.] You write the [artifact] and ask [the AI / your colleague] to identify out-of-date or incorrect information to correct.
    \item[5.] You write the [artifact] and ask [the AI / your colleague] to modify the organization or structure of the [artifact].
    \item[6.] You write notes and ideas and ask [the AI / your colleague] to synthesize the information into a cohesive [artifact].
    \item[7.] You write the [artifact] and ask [the AI / your colleague] to modify the tone and style to ensure it is suitable for the target audience.
    \item[8.] You write the [artifact] and ask [the AI / your colleague] to make spelling and grammar corrections.
    \item[9.] You write the [artifact] and ask [the AI / your colleague] to contribute new ideas.
    \item[10.] You write the [artifact] and ask [the AI / your colleague] to narrow the scope by removing less important ideas.
\end{itemize}

\subsubsection{Amount of contribution}
The following scenarios will focus on your perceptions of authorship for different amounts of contribution made by [the AI / your colleague]. For all scenarios, assume that contributions from [the AI / your colleague] are included in the final [artifact].

\begin{itemize}[leftmargin=0pt, itemindent=2em]
    \item[11.] You ask [an AI / your colleague] to write the full [artifact].
    \item[12.] You write a few sentences of the [artifact] and ask [an AI / your colleague] to write the rest.
    \item[13.] You write most of the [artifact] and ask [an AI / your colleague] to write a few sentences.
    \item[14.] You write the [artifact] on your own, without [AI / your colleague's] assistance.
    \item[15.] You write a few paragraphs of the [artifact] and ask [an AI / your colleague] to write a few other paragraphs.
\end{itemize}

\subsubsection{Initiative}
The following scenarios will focus on your perceptions of authorship when different parties initiate different types of contributions. For all scenarios, assume that contributions from [the AI / your colleague] are included in the final [artifact].

\begin{itemize}[leftmargin=0pt, itemindent=2em]
    \item[16.] {[The AI / Your colleague]} proactively suggests a complete [artifact] without being asked. The [artifact] is published without any modifications.
    \item[17.] You ask [the AI / your colleague] for ideas or feedback. You incorporate some of its recommendations into the [artifact].
    \item[18.] You ask [the AI / your colleague] to write the [artifact]. The [artifact] is published without any modifications.
    \item[19.] {[The AI / your colleague]} proactively provides ideas or feedback without being asked. You incorporate some of its recommendations into the [artifact].
\end{itemize}

\subsection{Attribution determination}

\begin{itemize}[leftmargin=0pt, itemindent=2em]
    \item[20.] \textit{Open response:} Please explain how you determined appropriate attribution for the scenarios in the previous three sections.
\end{itemize}

\begin{itemize}[leftmargin=0pt, itemindent=2em]
    \item[21.] Please indicate the importance of the following factors when you are determining who to attribute authorship to.
    \item Type of contribution (e.g. who wrote or modified wording vs. ideas)
    \item Amount of contribution (e.g. who wrote the most words)
    \item Initiative (e.g. who wrote the words or ideas first)
\end{itemize}

\textit{Participants rated the above items on the following scale.}

\begin{itemize}
    \item Not at all important
    \item Not very important
    \item Neither important nor unimportant
    \item Important
    \item Very important
\end{itemize}

\begin{itemize}[leftmargin=0pt, itemindent=2em]
    \item[22.] \textit{Open response:} Would any other factors affect your decision on who to attribute authorship to? If so, how would you describe the importance of these factors? 
\end{itemize}



    
     

\subsection{Closing}

\begin{itemize}[leftmargin=0pt, itemindent=2em]
    \item[23.] \textit{Open response:} Is there anything else you’d like to share on your perception of authorship in collaborative work?
\end{itemize}

\subsection{Demographic information}

Our research will investigate authorship in collaborative work between people and generative AI systems. As such, we’d like to understand more about your background with generative AI, in addition to a few standard demographic questions.

\begin{itemize}[leftmargin=0pt, itemindent=2em]
    \item[24.] In what ways do you use or work with generative AI applications (such as such as watsonx.ai, ChatGPT, DALL-E, Gemini, etc.)? Select all that apply.
    \item I use generative AI for school/academic purposes.
    \item I use generative AI for work purposes.
    \item I use generative AI for personal purposes (outside of my job or school).
    \item I develop, design, or study generative AI as part of my job.
    \item I am involved in training or tuning generative AI models.
\end{itemize}

\begin{itemize}[leftmargin=0pt, itemindent=2em]
    \item[25.] On average, how frequently do you use generative AI applications (such as watsonx.ai, ChatGPT, DALL-E, Gemini, etc.)?
    \item A couple times in my life
    \item Yearly
    \item Monthly
    \item Weekly
    \item Daily
\end{itemize}

\begin{itemize}[leftmargin=0pt, itemindent=2em]
    \item[26.] \textit{Optional:} What is your primary job category?
    \item Architect
    \item Communications
    \item Consultant
    \item Data Science
    \item Design
    \item Enterprise Operations
    \item Finance
    \item General Management
    \item Hardware Development \& Support
    \item Human Resources
    \item Information Technology \& Services
    \item Legal
    \item Manufacturing
    \item Marketing
    \item Marketing \& Communications
    \item Product Management
    \item Product Services
    \item Project Executive
    \item Project Management
    \item Research
    \item Sales
    \item Services Solutions Management
    \item Site Reliability Engineer
    \item Software Development \& Support
    \item Supply Chain
    \item Technical Services
    \item Technical Specialist
    \item Other: (write in)
\end{itemize}

\begin{itemize}[leftmargin=0pt, itemindent=2em]
    \item[27.] \textit{Optional:} What geography are you located in?
    \item Americas
    \item APAC
    \item EMEA
    \item Japan
\end{itemize}

\end{document}